\documentclass[twocolumn, tighten]{aastex63}
\usepackage{natbib}
\usepackage{multirow}

\emergencystretch=\maxdimen
\submitjournal{ApJ}
\shorttitle{Testing Rotating Regular Metrics as Candidates for Astrophysical Black Holes}
\shortauthors{R.~Kumar et al.}
\definecolor{MyDarkBlue}{rgb}{0,0.08,0.5}
\definecolor{MyDarkRed}{rgb}{0.7,0.02,0.02}
\definecolor{MyDarkGreen}{rgb}{0.0,0.7,0.0}

\usepackage{amsmath}

\begin{document}
\title{Testing Rotating Regular Metrics as Candidates for Astrophysical Black Holes}
	%
	\correspondingauthor{Rahul Kumar}
	\email{rahul.phy3@gmail.com}
	\author{Rahul Kumar}
	\affiliation{Centre for Theoretical Physics, Jamia Millia Islamia, New Delhi 110025, India}
	\author{Amit Kumar}
	\affiliation{Department of Applied Physics, Delhi Technological University, Delhi 110042, India}
	\author{Sushant G. Ghosh}
	\affiliation{Centre for Theoretical Physics, Jamia Millia Islamia, New Delhi 110025, India}
	\affiliation{Astrophysics and Cosmology Research Unit, School of Mathematics, Statistics and Computer Science, University of
		KwaZulu-Natal, Private Bag 54001, Durban 4000, South Africa}

\begin{abstract}
The Event Horizon Telescope, a global submillimeter wavelength very long baseline interferometry array, produced the first image of supermassive black hole M87* showing a ring of diameter $\theta_d= 42\pm 3\,\mu$as, inferred a black hole mass of $M=(6.5 \pm 0.7) \times 10^9 M_\odot $ and allowed us to investigate the nature of strong-field gravity. The observed image is consistent with the shadow of a Kerr black hole, which according to the Kerr hypothesis describes the background spacetimes of all astrophysical black holes. The hypothesis, a strong-field prediction of general relativity, may be violated in the modified theories of gravity that admit non-Kerr black holes. Here, we use the black hole shadow to investigate the constraints when rotating regular black holes (non-Kerr) can be considered as astrophysical black hole candidates, paying attention to three leading regular black hole models with additional parameters $g$ related to nonlinear electrodynamics charge. Our interesting results based on the systematic bias analysis are that rotating regular black holes shadows may or may not capture Kerr black hole shadows, depending on the values of the parameter $g$. Indeed, the shadows of Bardeen black holes ($g\lesssim 0.26 M$), Hayward black holes ($g\lesssim 0.65 M$), and non-singular black holes ($g\lesssim 0.25 M$) are indistinguishable from Kerr black hole shadows within the current observational uncertainties, and thereby they can be strong viable candidates for the astrophysical black holes. Whereas Bardeen black holes ( $g\leq 0.30182M$), Hayward black holes ($g\leq 0.73627M$), and non-singular black holes ($g\leq 0.30461M$), within the $1\sigma$ region for $\theta_d= 39\, \mu$as,  are consistent with the observed angular diameter of M87*.
\end{abstract}




\section{Introduction}
The Kerr hypothesis states that the astrophysical black hole candidates are well described by the Kerr metric (\cite{Kerr:1963ud}). Theoretically, this hypothesis is based on the uniqueness theorem that led to the general relativity's no-hair theorem stating that the Kerr (\citeyear{Kerr:1963ud}) and Kerr-Newman \citep{Newman:1965my} are the only stationary, axially symmetric, and asymptotically flat, respectively,  vacuum and electro-vacuum solutions of the Einstein equations  \citep{Israel:1967wq,Israel:1967za,Carter:1971zc,Hawking:1971vc,Robinson:1975bv}. But direct evidence of these solutions is still inconclusive and it may be difficult to rule out non-Kerr black holes \citep{Ryan:1995wh,Will:2005va,Bambi:2011jq}. With the current unprecedented observational techniques, it could become possible to probe deeper into the strong-field regime of gravity and eventually test the no-hair theorem.  A rotating regular black hole, a prototype non-Kerr metric, has an additional parameter $g$ due to magnetic charge arising from the nonlinear electrodynamics (NED) apart from mass $M$ and rotation parameter $a$ and encompasses the Kerr black hole ($g=0$) \citep{Kerr:1963ud}. Are such black holes candidates for the test of the no-hair theorem or the Kerr hypothesis?  The idea of a regular black hole was realized by the Bardeen (\citeyear{bardeen1968non}); based on the idea of Sakharov (\citeyear{Sakharov:1966aja}) and Gliner (\citeyear{Gliner}) he presented the first-ever static spherically symmetric regular black hole model. Since then, invoking a suitable NED Lagrangian density, a large number of regular black holes have been constructed \citep{Dymnikova:1992ux,Dymnikova:2004zc,AyonBeato:1998ub,AyonBeato:1999ec,Bronnikov:2000vy,Burinskii:2002pz,Berej:2006cc,Bronnikov:2005gm,Junior:2015fya,Sajadi:2017glu}, and more recently in Refs.~\citep{Fan:2016hvf,Bronnikov:2017tnz,Toshmatov:2018cks}.  Subsequently, several rotating regular black holes also have been introduced \citep{Bambi:2013ufa,Toshmatov:2014nya,Azreg-Ainou:2014pra,Ghosh:2014hea,Ghosh:2014pba,Rodrigues:2017tfm}. Regular black holes and their extensions have been extensively studied in the wide context of astrophysical phenomena, namely, particle motion \citep{Stuchlik:2014qja,Amir:2015pja,Garcia:2013zud,Amir:2016nti}, and gravitational lensing and shadows \citep{Eiroa:2013nra,Li:2013jra,Tsukamoto:2014tja,Schee:2015nua,Abdujabbarov:2016hnw,Amir:2016cen,Lamy:2018zvj,Jusufi:2018jof,Ovgun:2019wej,Kumar:2019pjp}. Interestingly, the size and shape of rotating regular black hole shadows are affected such that the shadow radius decreases and distortion increases monotonically with the parameters $g$, and thus a violation of the no-hair theorem significantly alters the shadow shape and size \citep{Bambi:2008jg,Johannsen:2010ru,Falcke:2013ola,Johannsen:2013rqa,Johannsen:2016uoh}. The black hole shadow relevance for testing the strong-field features of gravity, estimating black hole parameters and deducing any potential deviation from the Kerr geometry has resulted in comprehensive literature addressing shadows in both general relativity and modified gravities \citep{Hioki:2009na,Amarilla:2010zq,Amarilla:2011fx,Yumoto:2012kz,Atamurotov:2013sca,Grenzebach:2014fha,Ghasemi-Nodehi:2015raa,Abdujabbarov:2015rqa,Cunha:2016wzk,Amir:2017slq,Ayzenberg:2018jip,Perlick:2018iye,Wang:2018eui,Wang:2018prk,Shaikh:2019fpu,Mishra:2019trb,Long:2019nox,Konoplya:2019goy}.

Black hole shadows have become a physical reality with the recent detection of the horizon-scale image of the M87* black hole by the Event Horizon Telescope (EHT). Adopting a distance of $d=16.8$ Mpc and fitting the geometric crescent models obtained from the general-relativistic magnetohydrodynamic simulations to the M87* observational visibility data the mass of the M87* black hole is estimated $M=(6.5 \pm 0.7) \times 10^9 M_\odot $ \citep{Akiyama:2019cqa,Akiyama:2019fyp,Akiyama:2019eap}. The bright sharp photon ring, a projection along the null geodesics of the photons orbiting around the black hole, encompasses the shadow and explicitly depends on the black hole parameters while largely remaining independent of the detailed accretion models \citep{Beckwith:2004ae,Johannsen:2010ru,Johannsen:2015qca}. The constraints on the compact emission region size with angular diameter $\theta_d=42\pm 3\, \mu $as along with the central flux depression with a factor of $\geq 10$, which can be identified as the shadow, provide stringent evidence for the existence of the
black hole. Though the observed shadow of the M87* black hole is found to be consistent with that for the Kerr black hole as predicted by the general relativity, the current uncertainty in the measurement of spin angular momentum and the relative deviation of quadrupole moments do not eliminate non-Kerr black holes arising in modified gravities \citep{Akiyama:2019cqa,Akiyama:2019fyp,Akiyama:2019eap,Cardoso:2019rvt}.

Using the M87* black hole shadow, one can investigate the viability of different black hole models in explaining the observational data and put constraints on the black hole parameters. The main aim of this paper is to investigate if black hole shadow can help us to probe whether the rotating regular black holes can be a candidate for the astrophysical black holes. We made a systematic bias analysis between the model and injection shadows by considering the rotating regular black hole shadow as a model to fit with the Kerr shadow injections and for this purpose the shadow area $A$ and oblateness $D$ are used as observables to characterize the shadow \citep{Kumar:2018ple,Tsupko:2017rdo}. The best-fit values of the parameters ($a, g$) for which rotating regular black holes well capture the given Kerr shadow are determined. It is found that as the parameters lie in a certain special range, the rotating regular black holes shadows may illustrate the resemblance with the Kerr black hole shadows and are indiscernible within the current observational uncertainties and even can affirm the apparent asymmetry and angular size of the observed M87* shadow.
  
The paper is organized as follows: Section~\ref{sect2} is devoted to the brief review of the null geodesic equations in general stationary, axially symmetric regular black hole spacetimes and the recipe to investigate the shadow. In Section~\ref{sect3} we discuss the systematic bias analysis and check the compatibility of rotating regular black holes to explain the Kerr black hole shadows in Section~\ref{sect4}. In Section~\ref{sect5} we constrain the rotating regular black hole parameters with the aid of observed image of the M87* black hole. Finally, in Section~\ref{sect6} we summarize the obtained results and discuss the possibility of candidature of rotating regular black holes, with the shadow observations, for the astrophysical black holes.

\section{Rotating black hole shadow}\label{sect2}
A black hole surrounded by an optically thin emission region appears as a dark shadow in the observer's sky, which is an apparent cross-section of the gravitationally captured photon region confined by the innermost unstable photon orbits \citep{CT}. For this purpose, we study the photon geodesics around the rotating regular black hole spacetime, whose line element in Boyer$-$Lindquist coordinates ($t, r, \theta, \phi$) reads \citep{Bambi:2013ufa,Toshmatov:2014nya}
\begin{eqnarray}\label{rotmetric}
ds^2 & = & - \left( 1- \frac{2m(r)r}{\Sigma} \right) dt^2  - \frac{4am(r)r}{\Sigma  } \sin^2 \theta\, dt \, d\phi +
\frac{\Sigma}{\Delta}dr^2  \nonumber
\\ && + \Sigma\, d \theta^2+ \left[r^2+ a^2 +
\frac{2m(r) r a^2 }{\Sigma} \sin^2 \theta
\right] \sin^2 \theta\, d\phi^2,
\end{eqnarray}
and
\begin{eqnarray}
\Sigma = r^2 + a^2 \cos^2\theta,\;\;\;\;\;  \Delta = r^2 + a^2 - 2m(r)r.
\end{eqnarray}
where $m(r)$ is the mass function such that $\lim_{r\to\infty}m(r)=M$ and $a$ are the spin parameter defined as $a=J/M$; $J$ and $M$ are, respectively, the angular momentum and ADM mass of rotating black hole. The metric (\ref{rotmetric}) is Kerr (\citeyear{Kerr:1963ud}) and Kerr$-$Newman \citep{Newman:1965my} spacetimes, respectively, when $m(r)=M$ and $m(r)=M-Q^2/2r$. In general metric (\ref{rotmetric}) also describes various classes of rotating regular black holes when we choose $m(r)$ appropriately \citep{Ghosh:2014pba,Abdujabbarov:2016hnw,Amir:2016cen}. The photon motion in the spacetime (\ref{rotmetric}) is determined by the corresponding geodesics equations obtained from the Hamilton$-$Jacobi equation \citep{Carter:1968rr} 
\begin{eqnarray}
\label{HmaJam}
\frac{\partial S}{\partial \tau} = -\frac{1}{2}g^{\alpha\beta}\frac{\partial S}{\partial x^\alpha}\frac{\partial S}{\partial x^\beta} ,
\end{eqnarray}
where $\tau$ is the affine parameter along the geodesics, and $S$ is the Jacobi action, which reads as
\begin{eqnarray}
S=-{\cal E} t +{\cal L} \phi +S_r(r)+S_\theta(\theta) \label{action},
\end{eqnarray}
$S_r(r)$ and $S_{\theta}(\theta)$, respectively, are functions only of the $r$ and $\theta$ coordinates. The metric (\ref{rotmetric}) is time translational and rotational invariant, which leads to conserved quantities along geodesics, namely, energy $\mathcal{E}=-p_t$ and axial angular momentum $\mathcal{L}=p_{\phi}$, where $p_{\mu}$ is the photon's four-momentum. 
The Petrov-type $D$ character of metric~(\ref{rotmetric}) ensures the existence of Carter's separable constant $\mathcal{K}$, which eventually leads to the following complete set of null geodesics equations in the first-order differential form \citep{Carter:1968rr,Chandrasekhar:1985kt}:
\begin{eqnarray}
\Sigma \frac{dt}{d\tau}&=&\frac{r^2+a^2}{\Delta}\left({\cal E}(r^2+a^2)-a{\cal L}\right)  -a(a{\cal E}\sin^2\theta-{\mathcal {L}})\ ,\label{tuch}\\
\Sigma \frac{dr}{d\tau}&=&\pm\sqrt{\mathcal{V}_r(r)}\ ,\label{r}\\
\Sigma \frac{d\theta}{d\tau}&=&\pm\sqrt{\mathcal{V}_{\theta}(\theta)}\ ,\label{th}\\
\Sigma \frac{d\phi}{d\tau}&=&\frac{a}{\Delta}\left({\cal E}(r^2+a^2)-a{\cal L}\right)-\left(a{\cal E}-\frac{{\cal L}}{\sin^2\theta}\right)\ ,\label{phiuch}
\end{eqnarray}
where 
$\mathcal{V}_r(r)$ and $\mathcal{V}_{\theta}(\theta)$, respectively, are related to the effective potentials for radial and polar motion and are given by
\begin{eqnarray}\label{06}
\mathcal{V}_r(r)&=&\left((r^2+a^2){\cal E}-a{\cal L}\right)^2-\Delta({\cal K}+(a{\cal E}-{\cal L})^2)\label{rpot},\quad \\ 
\mathcal{V}_{\theta}(\theta)&=&{\cal K}-\left(\frac{{\cal L}^2}{\sin^2\theta}-a^2 {\cal E}^2\right)\cos^2\theta.\label{theta0}
\end{eqnarray}
The separability constant $\mathcal{K}$ is related to the Carter (\citeyear{Carter:1968rr}) constant of motion $\mathcal{Q}=\mathcal{K}+(a\mathcal{E}-\mathcal{L})^2$, which is essentially a manifestation of the isometry of metric (\ref{rotmetric}) along the second-order Killing tensor field. The black hole shadow boundary requires two dimensionless impact parameters \citep{Chandrasekhar:1985kt} 
\begin{equation}
\xi\equiv \mathcal{L}/\mathcal{E},\qquad \eta\equiv \mathcal{K}/\mathcal{E}^2,
\end{equation}
such that each geodesic is characterized by only these two parameters. Depending on the radial effective potential $\mathcal{V}_r(r)$ and the constants of motion, photons may follow scattering orbits, capturing orbits, and unstable orbits at the constant radii. These unstable orbits, constructing a photon region around the event horizon, delineate an apparent boundary separating the dark and bright regions on the observer's sky and account for the optical appearance of the black hole \citep{CT}. These orbits can be determined by the unique extremum of the potential in the exterior region to the event horizon $r_+$ i.e., $r>r_+$, such that
\begin{equation}
\left.\mathcal{V}_r\right|_{(r=r_p)}=\left.\frac{\partial \mathcal{V}_r}{\partial r}\right|_{(r=r_p)}=0,\,\, \text{and}\,\, \left.\frac{\partial^2 \mathcal{V}_r}{\partial r^2}\right|_{(r=r_p)}> 0,\label{vr} 
\end{equation}
where $r_p$ is the unstable photon orbit radius.  Solving Eq.~(\ref{vr}) for Eq.~(\ref{rpot}) results in the critical impact parameters ($\xi_c, \eta_c$) for the unstable orbits \citep{Abdujabbarov:2016hnw,Kumar:2018ple}

\begin{align}
\xi_c=&\frac{[a^2 - 3 r_p^2] m(r_p) + r_p [a^2 + r_p^2] [1 + m'(r_p)]}{a [m(r_p) + r_p [-1 + m'(r_p)]]},\nonumber\\
\eta_c=&-\frac{r_p^3}{a^2 [m(r_p) + r_p [-1 + m'(r_p)]]^2}\Big[r_p^3 + 9 r_p m(r_p)^2 \nonumber\\
&+ 2 [2 a^2 + r_p^2+r_p^2 m'(r_p)] r_pm'(r_p) \nonumber\\
&-  2 m(r_p) [2 a^2 + 3 r_p^2 + 3 r_p^2 m'(r_p)]\Big],\label{impactparameter}
\end{align}

where $'$ stands for the derivative with respect to the radial coordinate. If we consider the Kerr black hole ($m(r)=M$), Eq.~(\ref{impactparameter}) reduces to \citep{Chandrasekhar:1985kt}
\begin{align}
\xi_c=&\frac{M(r_p^2-a^2)-r_p(r_p^2+a^2-2Mr_p)}{a(r_p-M)},\nonumber\\
\eta_c=&-\frac{r_p^3\left[4Ma^2-r_p(r_p-3M)^2\right]}{a^2(r_p-M)^2}.\label{impactparameterkerr}
\end{align}

In particular, photons with $\mathcal{\eta}_c=0$ form planar circular orbits confined only to the equatorial plane, whereas $\mathcal{\eta}_c>0$ lead to three-dimensional spherical orbits \citep{Chandrasekhar:1985kt}. In rotating spacetimes, photons can either have prograde motion or retrograde motion, whose respective radii at the equatorial plane, $r_p^{-}$ and $r_p^{+}$, can be identified as the real positive roots of $\eta_c=0$ for $r_p\geq r_+$, and all other spherical photon orbits have radii $r_p^-< r_p<r_p^+$. Further, the maximum latitude of spherical orbits depends on the angular momentum of photons, viz., the smaller the angular momentum the larger the orbit latitude. Photons must have zero angular momentum to reach the polar plane of the black hole, whose orbit radius $r_p^0$, such that $r_p^-\leq r_p^0\leq r_p^+$, can be determined by the zeros of $\xi_c=0$. For the Kerr black hole, these photon orbit radii $r_p$ are \citep{Teo:2003}

\begin{align}
r_p^-&=2M\left[1+ \cos\left(\frac{2}{3}\cos^{-1}\left[-\frac{|a|}{M}\right]\right) \right],\nonumber\\
r_p^+&=2M\left[1+ \cos\left(\frac{2}{3}\cos^{-1}\left[\frac{|a|}{M}\right]\right) \right],\nonumber\\
r_p^0&=M+2\sqrt{M^2-\frac{1}{3}a^2}\,\cos\Big[\frac{1}{3}\cos^{-1}\Big(\frac{M(M^2-a^2)}{(M^2-\frac{1}{3}a^2)^{3/2}}\Big)\Big],
\end{align}  
which for the Schwarzschild black hole ($a=0$) takes the degenerate value $r_p^-=r_p^0=r_p^+=3M$ and for the extremal Kerr black hole ($a=M$) takes the values $r_p^-=M$, $r_p^0=(1+\sqrt{2})M$ and $r_p^+=4M$. 
\begin{figure}
	\begin{tabular}{c c}
	\includegraphics[scale=0.4]{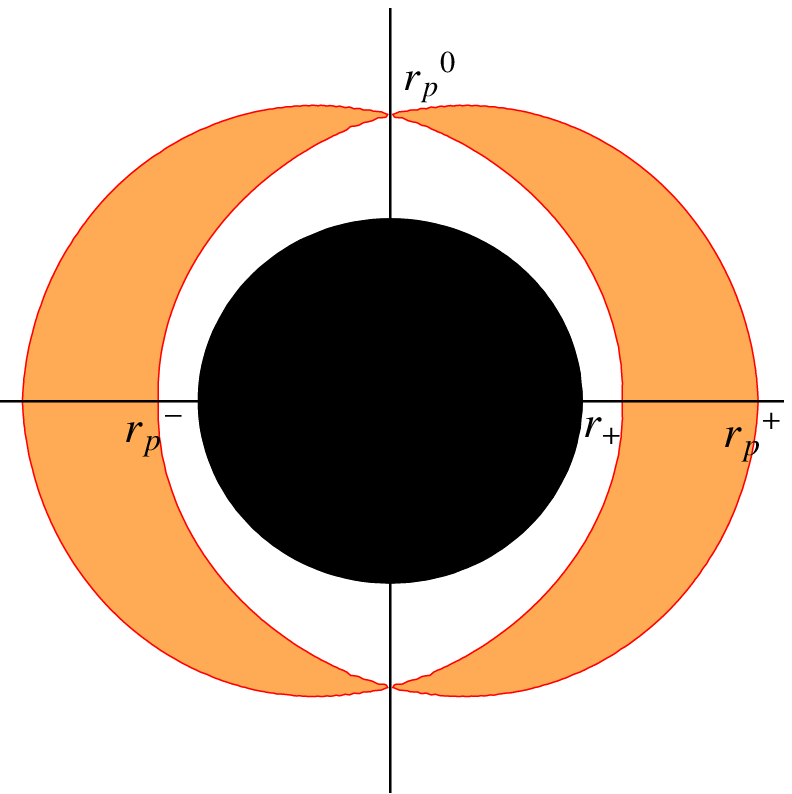}&
	\includegraphics[scale=0.47]{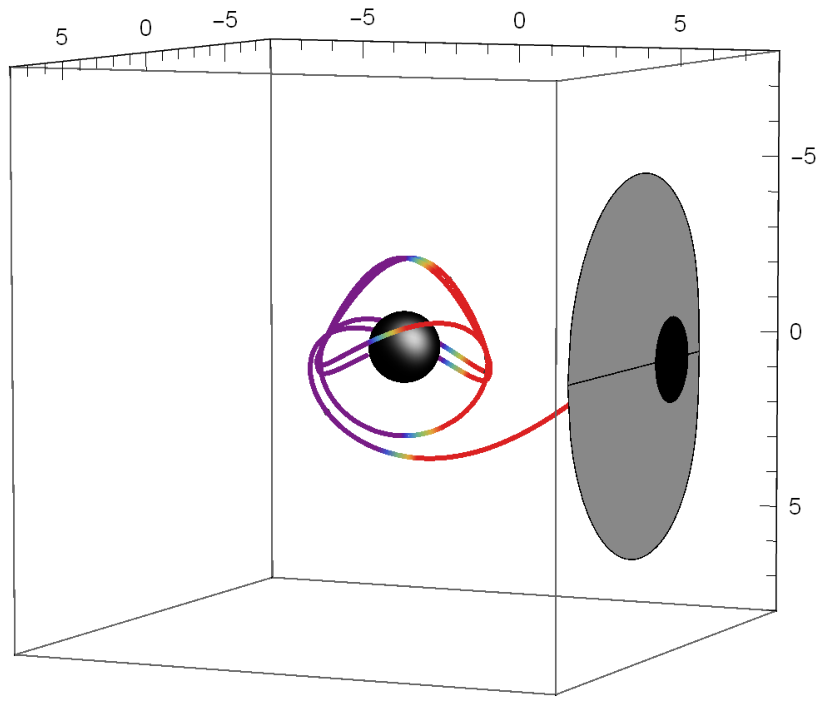}
	\end{tabular}
\caption{(Left panel) Photon region (shaded orange region) around a Kerr black hole (black disk). (Right panel) Photon orbit projection as a black hole shadow on the observer's image plane. In the projected image, the small dark black disk and gray deformed region, respectively, correspond to the black hole horizon and the shadow.}\label{Orbit}
\end{figure}
Thus the gravitationally lensed image of the photon region around the black hole yields the apparent shadow, whose boundary can be traced by the loci of critical impact parameters ($\xi_c, \eta_c$). For an observer at the position ($r_o,\theta_o$), in the far exterior region of the black hole, the shadow boundary can be described in terms of the celestial coordinates \citep{bardeen1973,CT}
\begin{equation}
\alpha=-r_o\frac{p^{(\phi)}}{p^{(t)}},\qquad \beta=r_o\frac{p^{(\theta)}}{p^{(t)}},
\end{equation} 
where $p^{(\mu)}$ is the photon four-momentum measured on an orthonormal-tetrad basis. On using geodesic Eqs.~(\ref{tuch}), (\ref{th}), and (\ref{phiuch}), the celestial coordinates yield
\begin{eqnarray}
&&\alpha=-\left. r_o\frac{\xi_c}{\sqrt{g_{\phi\phi}}(\zeta-\gamma\xi_c)}\right|_{(r_o,\theta_o)},\nonumber\\
&&\beta=\pm\left. r_o\frac{\sqrt{\mathcal{V}_{\theta}(\theta)}
}{\sqrt{g_{\theta\theta}}(\zeta-\gamma\xi_c)}\right|_{(r_o,\theta_o)},~\label{Celestial}
\end{eqnarray} 
where 
\begin{eqnarray}
\zeta=\sqrt{\frac{g_{\phi\phi}}{g_{t\phi}^2-g_{tt}g_{\phi\phi}}},\qquad \gamma=-\frac{g_{t\phi}}{g_{\phi\phi}}\zeta.
\end{eqnarray}
The ($\alpha,\beta$) in Eq.~(\ref{Celestial}), respectively, denote the apparent displacement along the perpendicular and parallel axes to the projected axis of the black hole symmetry. For an observer sitting in the asymptotically flat region ($r_o\to\infty$), the celestial coordinates Eq.~(\ref{Celestial}) can be simplified as \citep{bardeen1973}
\begin{equation}
\alpha=-\xi_c\csc\theta_o,\qquad \beta=\pm\sqrt{\eta_c+a^2\cos^2\theta_o-\xi_c^2\cot^2\theta_o}.\label{pt}
\end{equation} 
The celestial coordinates have the exact same form as that for a Kerr black hole but with different $\eta_c$ and $\xi_c$. The photon region and its projected image on the observer's plane for a Kerr black hole are shown in Fig.~\ref{Orbit}. The parametric plot of Eqs.~(\ref{Celestial}) or (\ref{pt}) in the ($\alpha,\beta$) plane can cast a variety of black hole shadows for different choices of black hole mass function $m(r)$ \citep{Abdujabbarov:2016hnw,Amir:2016cen,Kumar:2019pjp}. Therefore, it is pertinent to check whether shadow observations can be used as a tool to distinguish non-Kerr (rotating regular) black holes from the Kerr black hole or can place constraints on the deviation parameters. Despite that the rotating regular black holes are considerably  different from the Kerr black hole, it may be difficult, at least in some cases, to differentiate some rotating regular black holes (e.g., rotating Bardeen black hole) from a Kerr black hole \citep{Li:2013jra,Tsukamoto:2014tja}. In what follows, we explore the possibilities of rotating regular black holes as astrophysical black hole candidates from the comparability of their shadows. 
\section{Systematic bias analysis}\label{sect3}
We consider the Kerr black hole shadows as the injection and simulated shadows of rotating regular black holes as models to fit with the injection. For a fixed observer position ($r_o, \theta_o$) and a given black hole mass $M$, the injection solely depends on the black hole spin parameter $a$, whereas model shadows depend on both spin $a$ and deviation parameter $g$. This deviation parameter can significantly alter the shape and size of shadows when compared with the Kerr black hole shadow  \citep{Abdujabbarov:2016hnw,Amir:2016cen,Kumar:2019pjp}. Consider two astronomical observables, namely, area $A$ and oblateness $D$, respectively, given by  \citep{Kumar:2018ple,Tsupko:2017rdo}
\begin{equation}
A=2\int{\beta(r_p) d\alpha(r_p)}=2\int_{r_p^{-}}^{r_p^+}\left( \beta(r_p) \frac{d\alpha(r_p)}{dr_p}\right)dr_p,\label{Area}
\end{equation} 
\begin{equation}
D=\frac{\alpha_r-\alpha_l}{\beta_t-\beta_b}.\label{Oblateness}
\end{equation}
Here, $A$ and $D$, respectively, characterize the size and shape of shadow; a single observable induces degeneracy between the spin parameter and the deviation parameter \citep{Tsukamoto:2014tja}. Oblateness parameter $D$ is defined as the ratio of horizontal and vertical diameters of the shadow as subscripts $r, l, t$, and $b$, respectively, standing for the right, left, top, and bottom of the shadow silhouette such that $D\neq 1$ refers to the rotating black hole \citep{Kumar:2018ple,Tsupko:2017rdo}. For a given black hole model, these observables uniquely characterize the shadow over the black hole parameter space ($a,g$). For a fixed spin parameter, these observables can take considerably different values for model and injection shadows \citep{Abdujabbarov:2016hnw,Amir:2016cen,Kumar:2019pjp}. It is also likely that a given set of shadow observables can correspond to more than one model shadows, though the values of black hole parameters are different \citep{Kumar:2018ple}. To explore this avenue, we make a systematic bias analysis by minimizing the $\chi ^2_{\text{red}}$ between the model and the injection to check their compatibility and to constrain the model parameter space ($a,g$). For fixed values of free parameters mass $M$, distance $d$, and inclination angle $\theta_o$, we define the $\chi ^2_{\text{red}} $ as a function of the spin and deviation parameters
\begin{equation}
\chi ^2_{\text{red}}(a,g,a_K)=\frac{1}{2}\sum ^2_{i=1}\bigg[\frac{\alpha ^i(a,g)- \alpha ^i_K(a_K)}{\sigma_i} \bigg]^2.
\end{equation}
Here, $\alpha ^i_K$ and $\alpha ^i$ are, respectively, the injection and model shadow observables $[D, A]$, where $a_K$ and $a$ are the injection and model spin parameters. In this model fitting procedure, the nonzero values of  $\chi ^2_{\text{red}}$ quantify the departure of rotating regular black hole shadows from that of Kerr black hole shadows, such that for  $\chi ^2_{\text{red}}\leq 1$ the rotating regular black hole shadow can capture the Kerr shadow and both are indistinguishable within the current observational uncertainties. Whereas if $\chi ^2_{\text{red}} >1$, then the current shadow observations can easily discern two black holes and constraints can be placed on regular black hole parameters. For the fixed values of the injected spin parameter $a_K$ and the deviation parameter $g$, the best-fit values of the model spin $a$ by minimizing the $\chi ^2_{\text{red}}$ between the injection and model shadows are obtained. It can be expected without a priori that for the large values of the deviation parameter $g$, the extracted best-fit values of the model spin $a$ will significantly deviate from the injected spins $a_K$. The constraints on the model parameter space ($a,g$) are obtained for which model shadows at best resemble the injected shadows. The standard deviation $\sigma_i$ is assumed to be $10\%$ of the range of each observable, which is the current uncertainty in the observational measurements of the EHT \citep{Akiyama:2019cqa,Akiyama:2019fyp,Akiyama:2019eap}.

\section{Application to regular black holes}\label{sect4}
The systematic bias analysis, introduced in the previous section, allows us to analyze whether the deviations of rotating regular black holes shadows from the Kerr shadows are large enough to be detectable with the current black hole shadow observations.  We examine three well-known rotating regular black holes, viz., Bardeen, Hayward, and non-singular black holes. The shadows of these regular models have received significant attention and their shape and size are considerably different from those of the Kerr black hole shadows \citep{Abdujabbarov:2016hnw,Amir:2016cen,Kumar:2019pjp}. Henceforth, for our purposes, we assume that the inclination angle is $\theta_o=\pi/2$ and consider the rotating regular black hole shadow as a model to fit with the Kerr black hole shadow injection.
\subsection{Bardeen black holes}
The Bardeen (\citeyear{bardeen1968non}) black hole metric is asymptotically ($r\to\infty$) flat, and near origin ($r\to 0$) behaves as the de-Sitter. The rotating regular Bardeen black hole \citep{Bambi:2013ufa,Toshmatov:2014nya} is described by metric (\ref{rotmetric}) with the mass function \citep{bardeen1968non} 
\begin{equation}
m(r)=M\left(\frac{r^2}{r^2 + g^2}\right)^{3/2}, \label{Bardeenmass}
\end{equation}
where the deviation parameter $g$ can be identified as the magnetic monopole charge \citep{AyonBeato:2000zs}. The rotating Bardeen black hole metric has been tested with the X-ray data from the disk around the black hole candidate in the Cygnus X-1 ($M=14.8\, M_{\odot},\, d=1.86\, \text{kpc},\, \theta_o=27.1 ^{o}$) \citep{Bambi:2014nta}. Such that the $3\sigma$ bounds $a_K>0.95 M$ \citep{Gou:2011nq} and $a_K>0.983M$ \citep{Gou:2013dna} for the Kerr metric infer a bound on Bardeen black hole parameters, respectively, $a>0.78M$ and $g<0.41M$, and $a>0.89M$ and $g<0.28M$ \citep{Bambi:2014nta}.
\begin{figure*}
\begin{center}

	\begin{tabular}{c c}
	\includegraphics[scale=0.73]{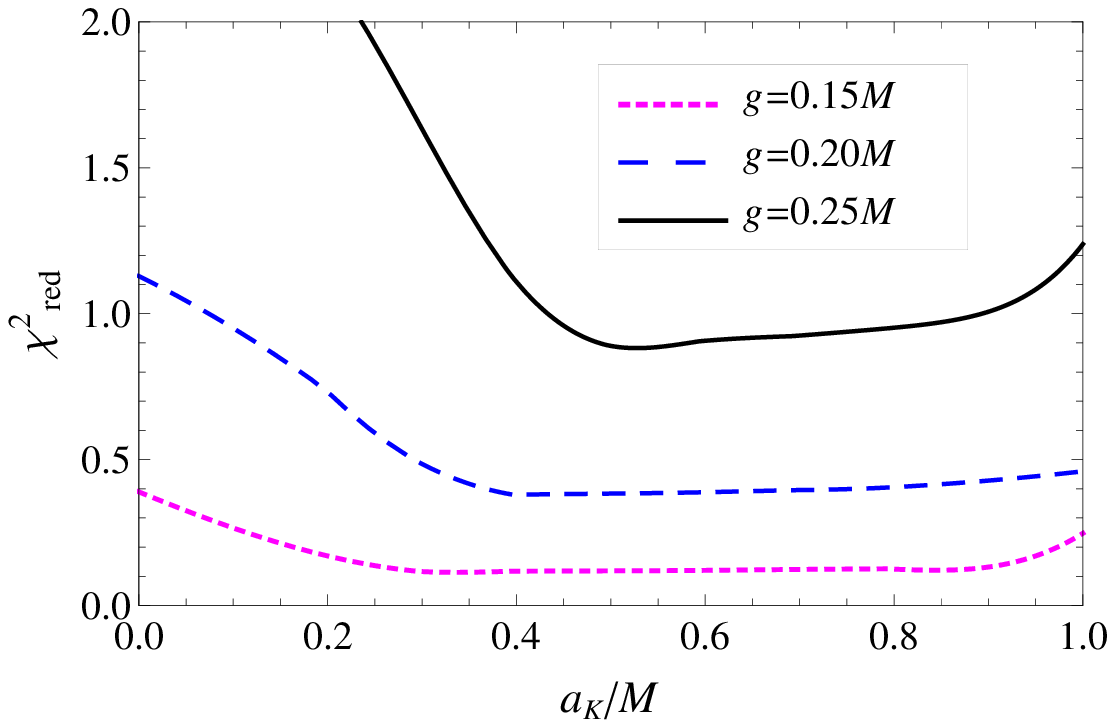} &
	\includegraphics[scale=0.73]{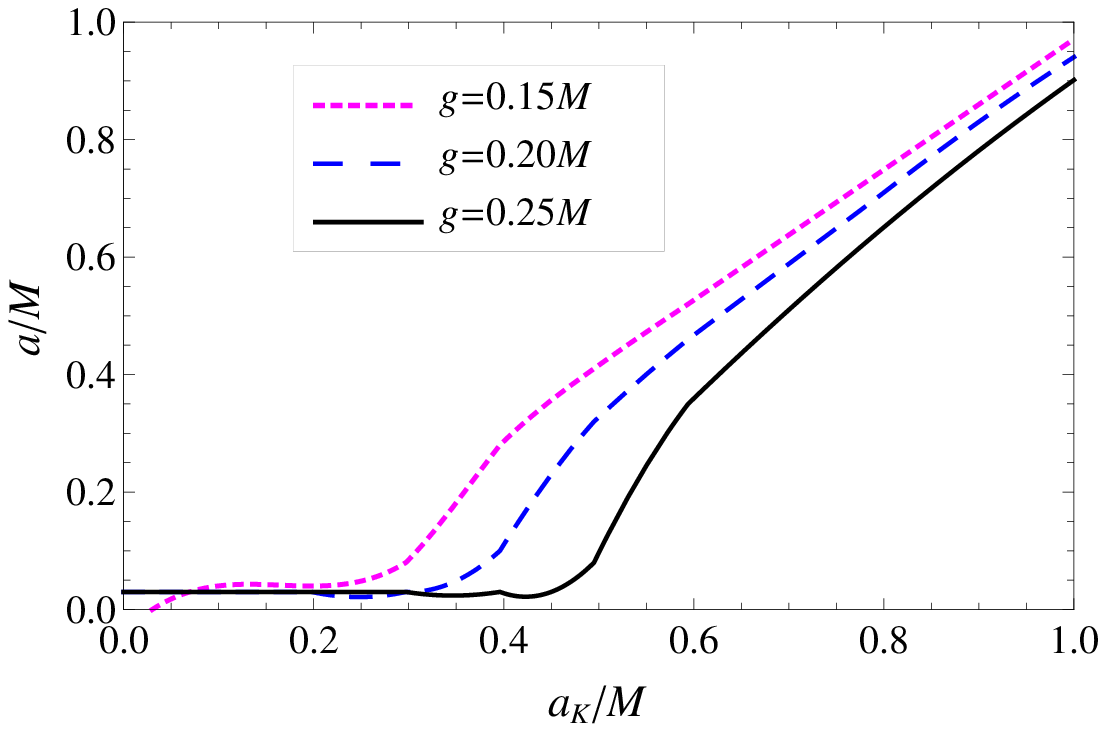}
	\end{tabular}
	\caption{Minimized $\chi ^2_{\text{red}}$ (left panel) and extracted values of the rotating Bardeen black hole spin parameter $a$ (right panel) for the best-fit model shadow as a function of the injected Kerr spin $a_K$.}   \label{BardeenFig}
	\begin{tabular}{c c c}
		\includegraphics[scale=0.74]{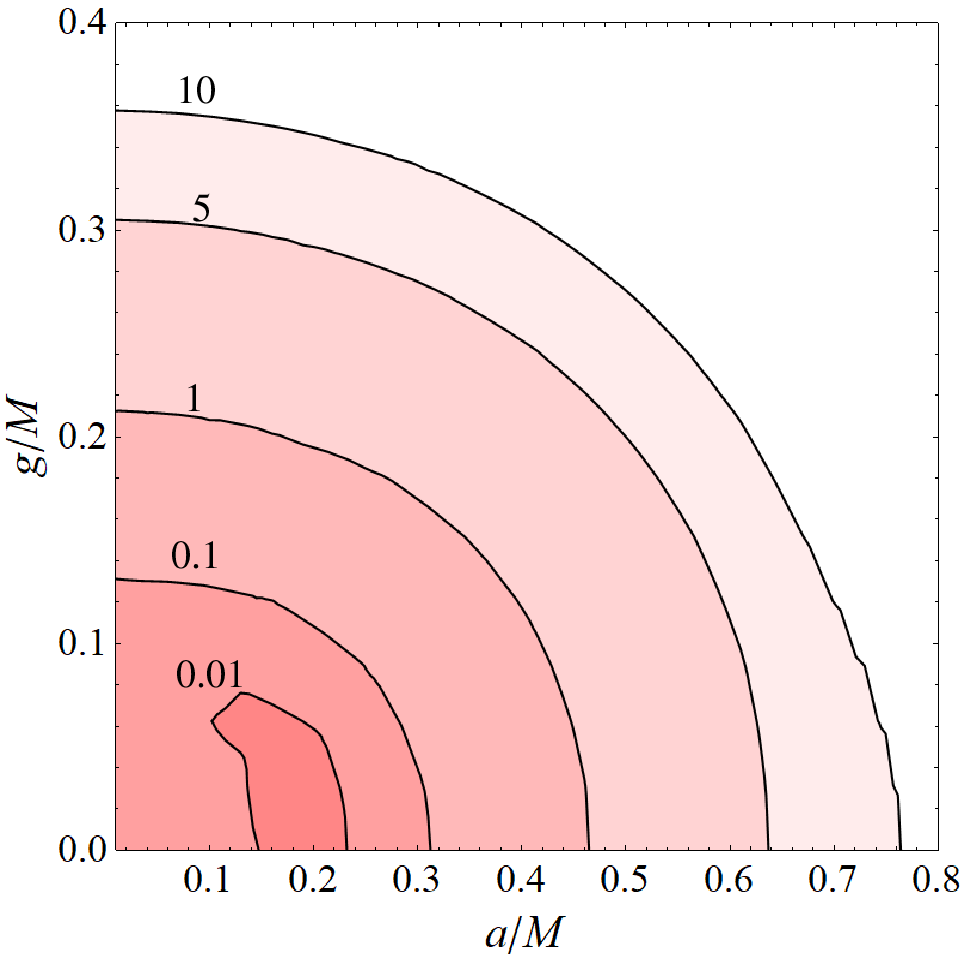} &
		\includegraphics[scale=0.74]{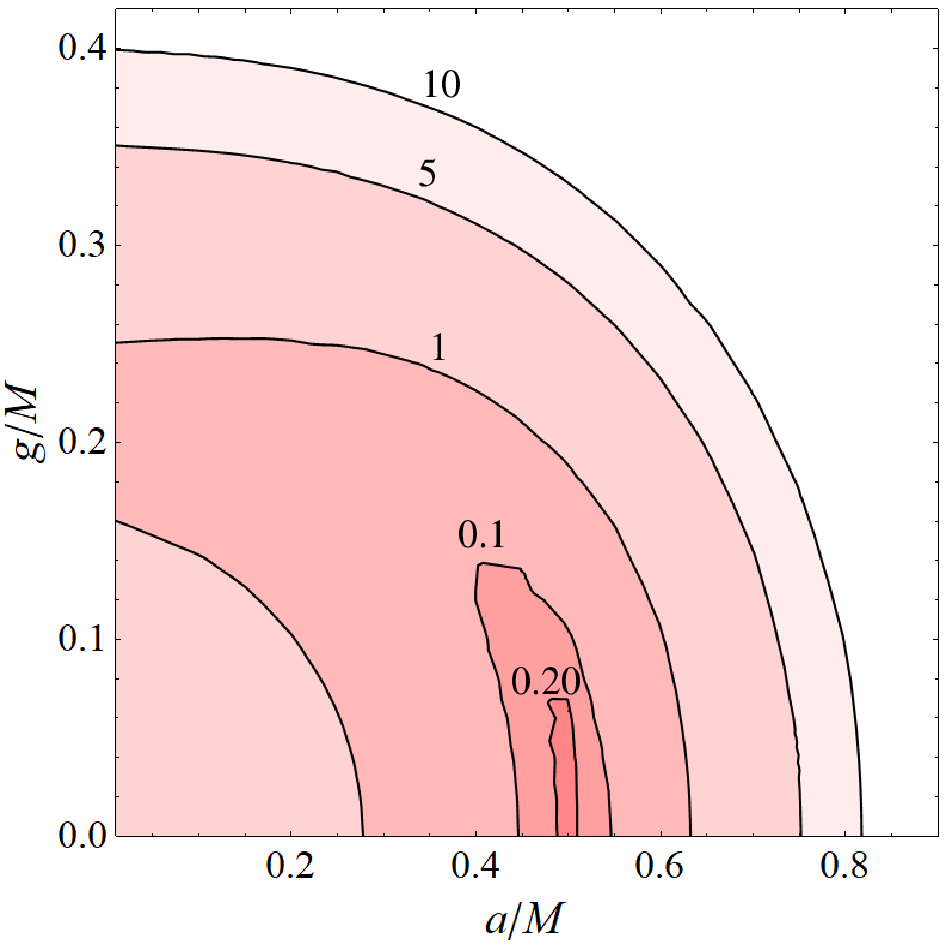}\\
		\includegraphics[scale=0.74]{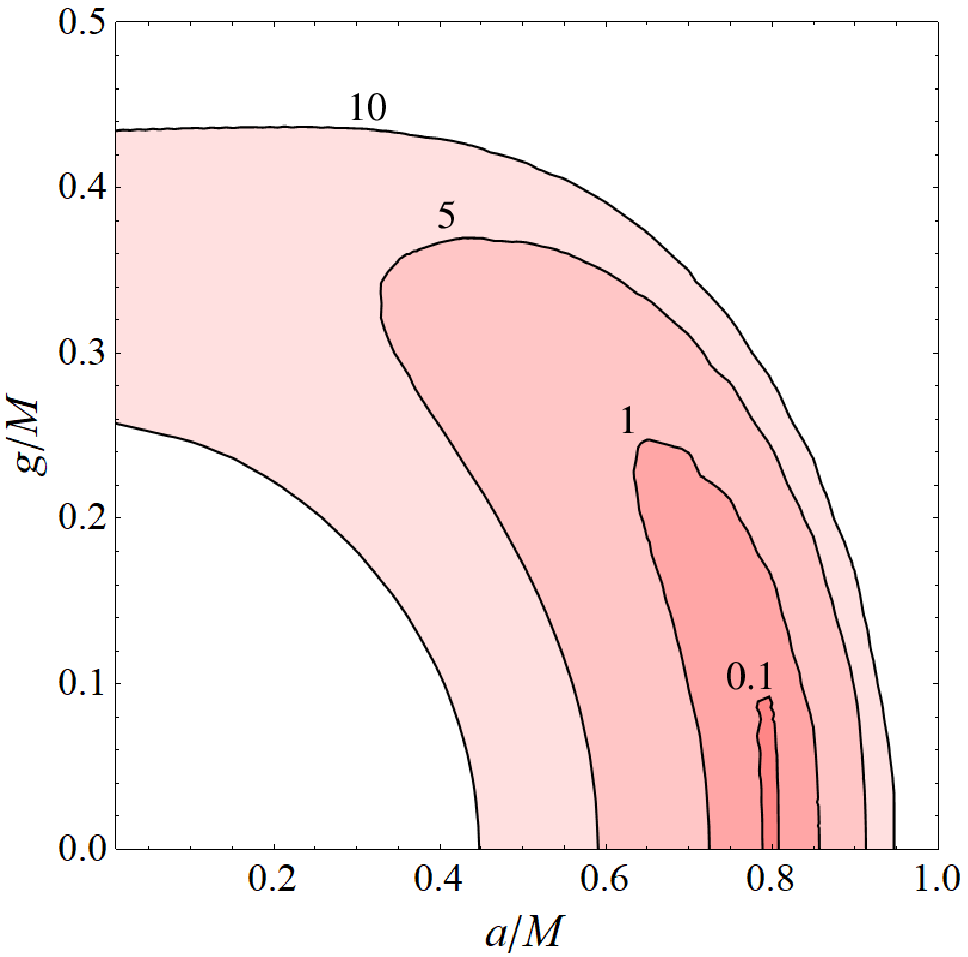}&
		\includegraphics[scale=0.74]{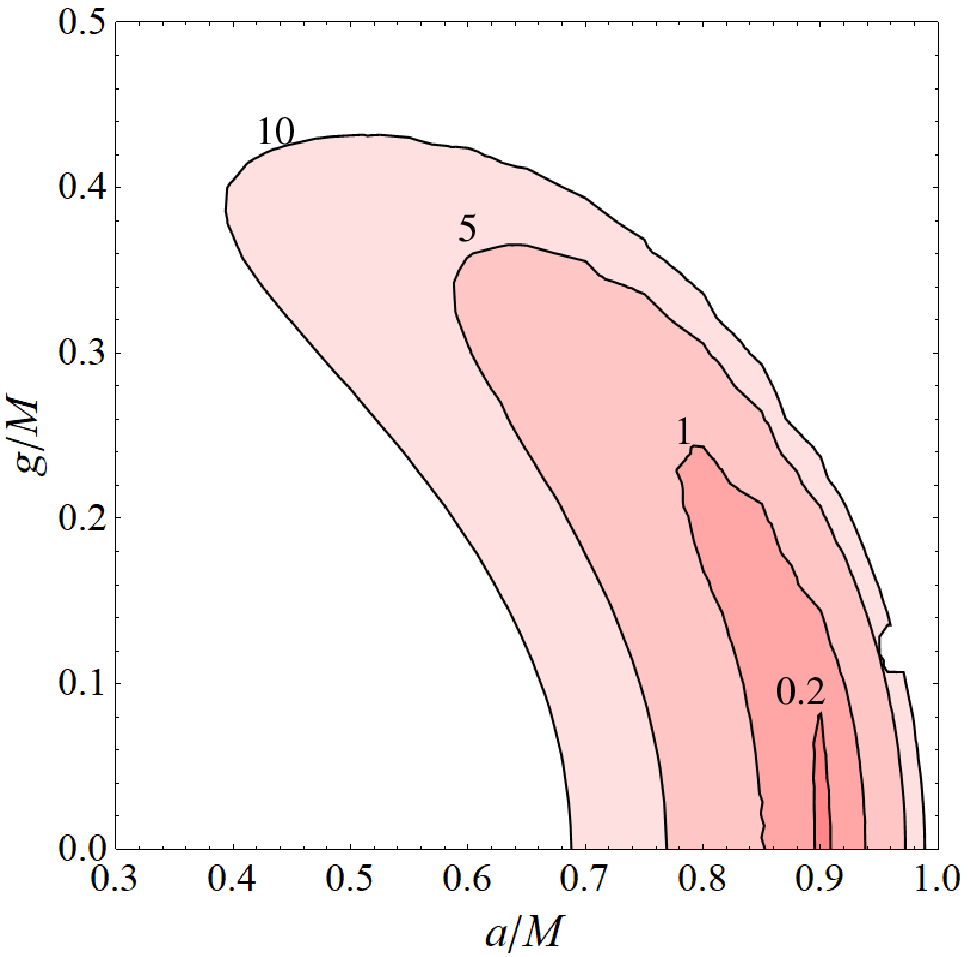}
	\end{tabular}
	\caption{$\chi ^2_{\text{red}}$ between rotating Bardeen black hole and Kerr black hole shadows as a function of ($a,g$) for $a_K=0.20M$, $0.50M$, $0.80M$ and $0.90M$ .}   \label{BardeenFig1}
\end{center}
\end{figure*}
We begin with the calculations of the observables $A$ and $D$ for the rotating Bardeen black hole (model) and the Kerr black hole (injection) shadows and investigate whether given shadow observables for the injection shadow can also correspond to the modeled shadow. The model shadow area is found to decrease with both $a$ and $g$, such that for $a=a_K$, the model shadow area is always smaller than that for the injection shadow area \citep{Kumar:2018ple}. It also turns out that the oblateness $D$ decreases with both $a$ and $g$ \citep{Kumar:2018ple}. The behavior of $\chi ^2_{\text{red}}$ between the model and injection shadows with varying injected spin $a_K$ is depicted in Fig.~\ref{BardeenFig}. It is evident that $\chi ^2_{\text{red}}$ shows a non-monotonic behavior with increasing injected spin, such that $\chi ^2_{\text{red}}$ is relatively large for the small values of $a_K$ then decreases with increasing $a_K$ and again increases for the near-extremal value of $a_K$. For sufficiently smaller values of $g$, $\chi ^2_{\text{red}}$ is less than $1$ for almost all values of injected spin $a_K$ (see Fig.~\ref{BardeenFig}). Whereas for higher values of $g$, model shadows could not substantially capture the injected shadows, as the $\chi ^2_{\text{red}}>1$ for all $a_K$. This placed a bound on model parameter $g$, such that for $g\lesssim 0.26 M$ model shadows are indistinguishable from the injected shadows. The best-fit values of the model spin parameter $a$ for different values of $g$ are extracted, which as a function of injected spin $a_K$ are shown in Fig.~\ref{BardeenFig}. The extracted model spin is biased from the injected spin and this disparity further increases with increasing $g$. The contour plots of $\chi ^2_{\text{red}}$ as a function of ($a,g$) for various values of $a_K$ are shown in Fig.~\ref{BardeenFig1}. One can deduce that for the finite parameter space ($a,g$), model shadows are well consistent with the injected shadows within the current observational uncertainties. Figure \ref{BardeenFig1} suggests that the parameter space viable for the indistinguishability of two shadows is bounded by $\chi ^2_{\text{red}}\leq 1$ and decreases with increasing $a_K$, viz., constraints on rotating Bardeen black hole parameters are substantially stronger for rapidly rotating Kerr black hole injections. 

\subsection{Hayward black holes}
Another thoroughly studied regular black hole model was proposed by Hayward (\citeyear{Hayward:2005gi}). Besides the mass $M$ it has one additional parameter $\ell$, which determines the length associated with the region concentrating the central energy density, such that modifications in the spacetime metric appear when the curvature scalar becomes comparable with $\ell^{-2}$. The spherically symmetric Hayward black hole model is identified as an exact solution of the general relativity minimally coupled to NED with magnetic charge $g$, where $g$ is related to $\ell$ via $g^3=2M\ell^2$ \citep{Fan:2016hvf}.
The rotating Hayward black hole is also Kerr-like black hole described by the metric (\ref{rotmetric}) with the mass function \citep{Hayward:2005gi,Bambi:2013ufa}
\begin{equation}
m(r)=\frac{Mr^3}{r^3+g^3}.\label{Haywardmass}
\end{equation}
The photon region and shadows of rotating Hayward black holes have been extensively discussed \citep{Abdujabbarov:2016hnw, Kumar:2019pjp,Liu:2017ifc}.
\begin{figure*}
\begin{center}	
	\centering
	\begin{tabular}{c c}
		\includegraphics[scale=0.73]{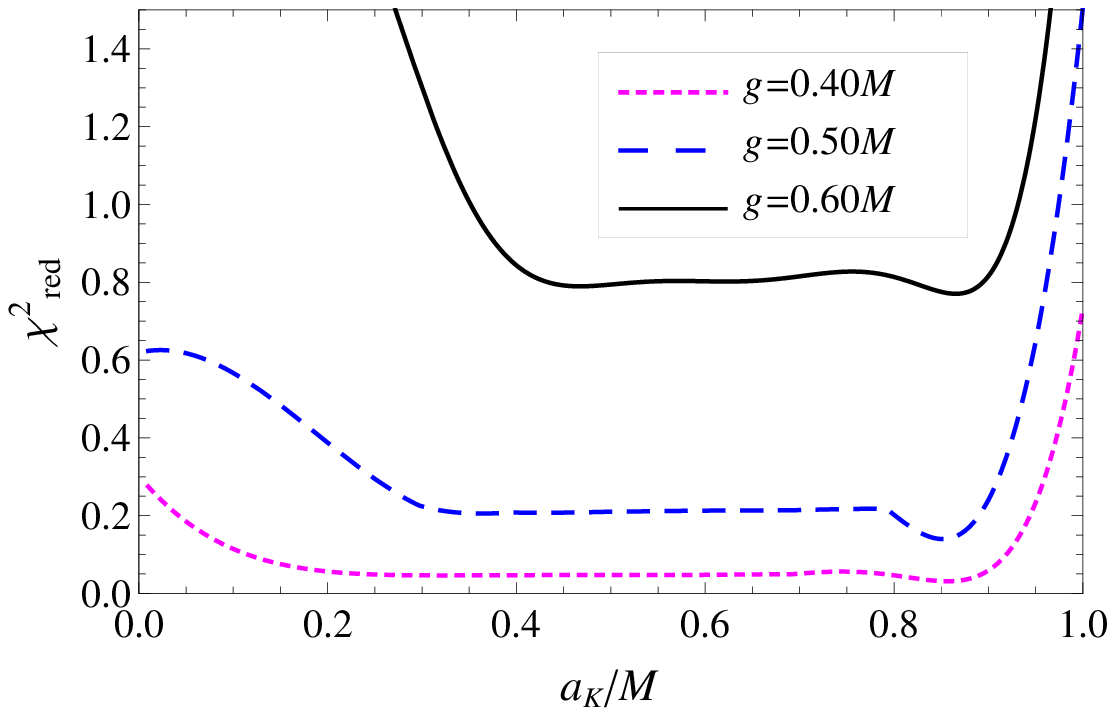} &
		\includegraphics[scale=0.73]{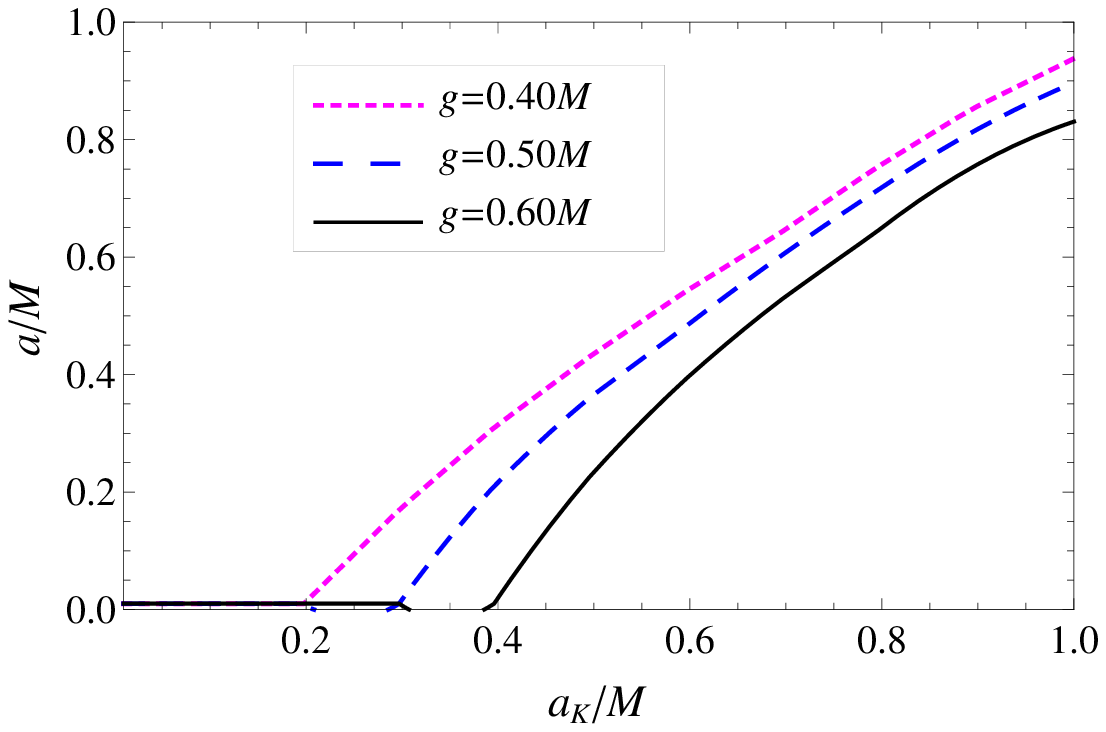}
	\end{tabular}
	\caption{Minimized $\chi ^2_{\text{red}}$ (left panel) and extracted values of the rotating Hayward black hole spin parameter $a$ (right panel) for the best-fit model shadow as a function of the Kerr spin $a_K$.} \label{HaywardFig}
	\begin{tabular}{c c }
		\includegraphics[scale=0.75]{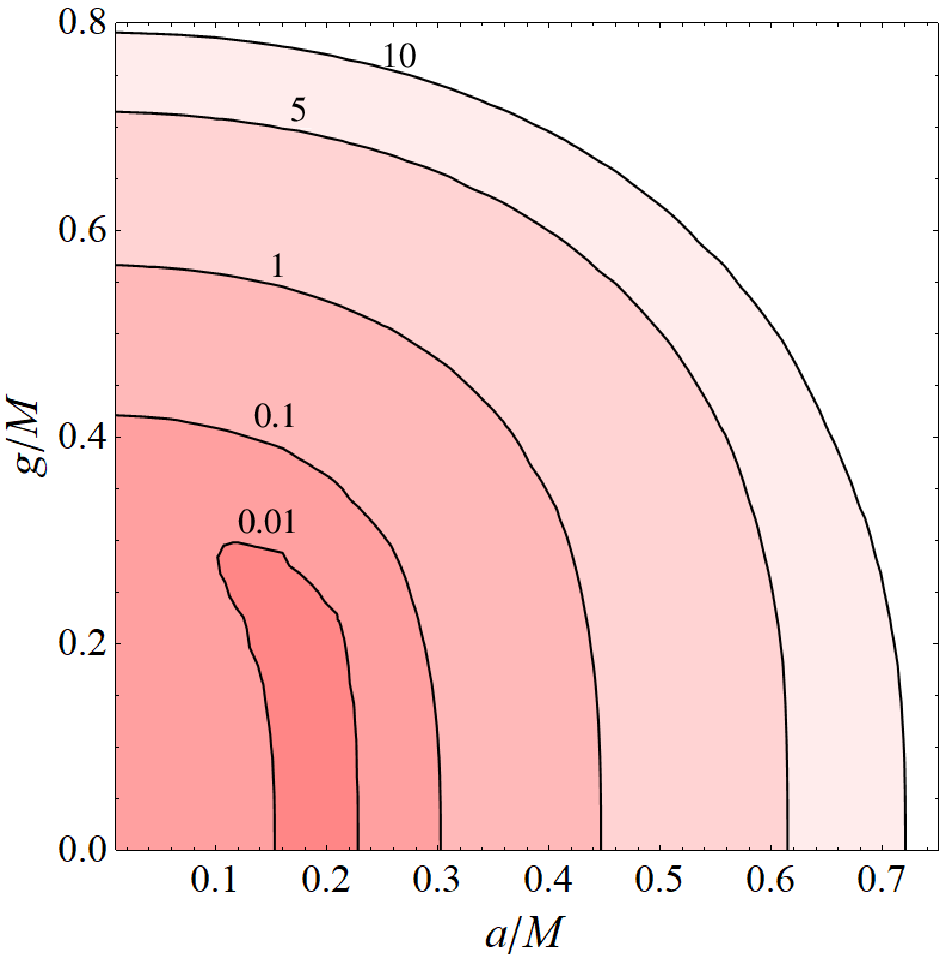} &
		\includegraphics[scale=0.75]{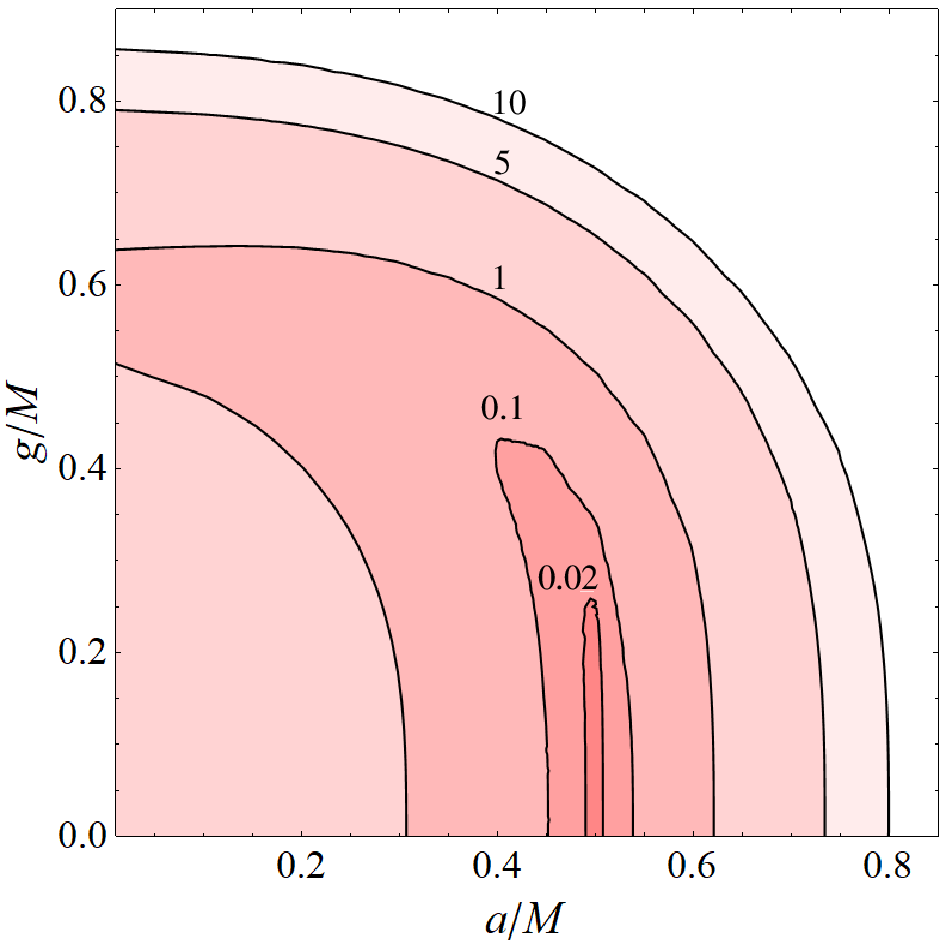}\\	\includegraphics[scale=0.75]{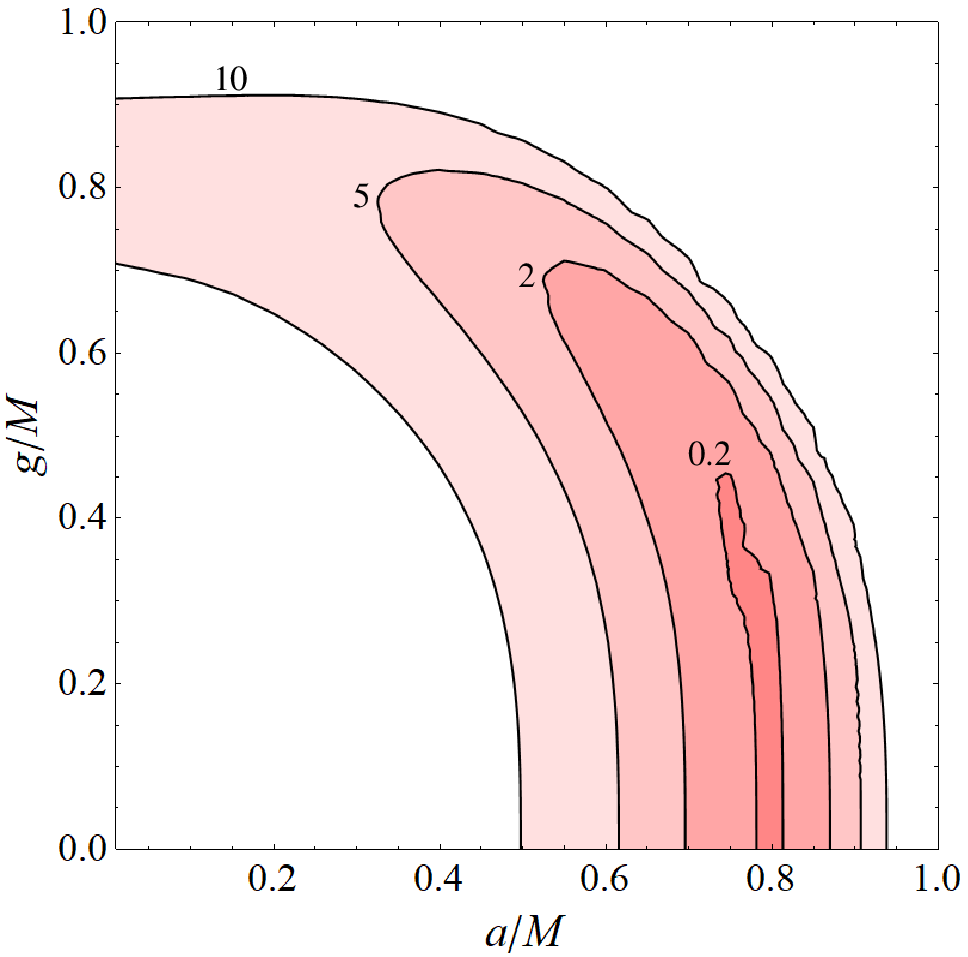}&
		\includegraphics[scale=0.75]{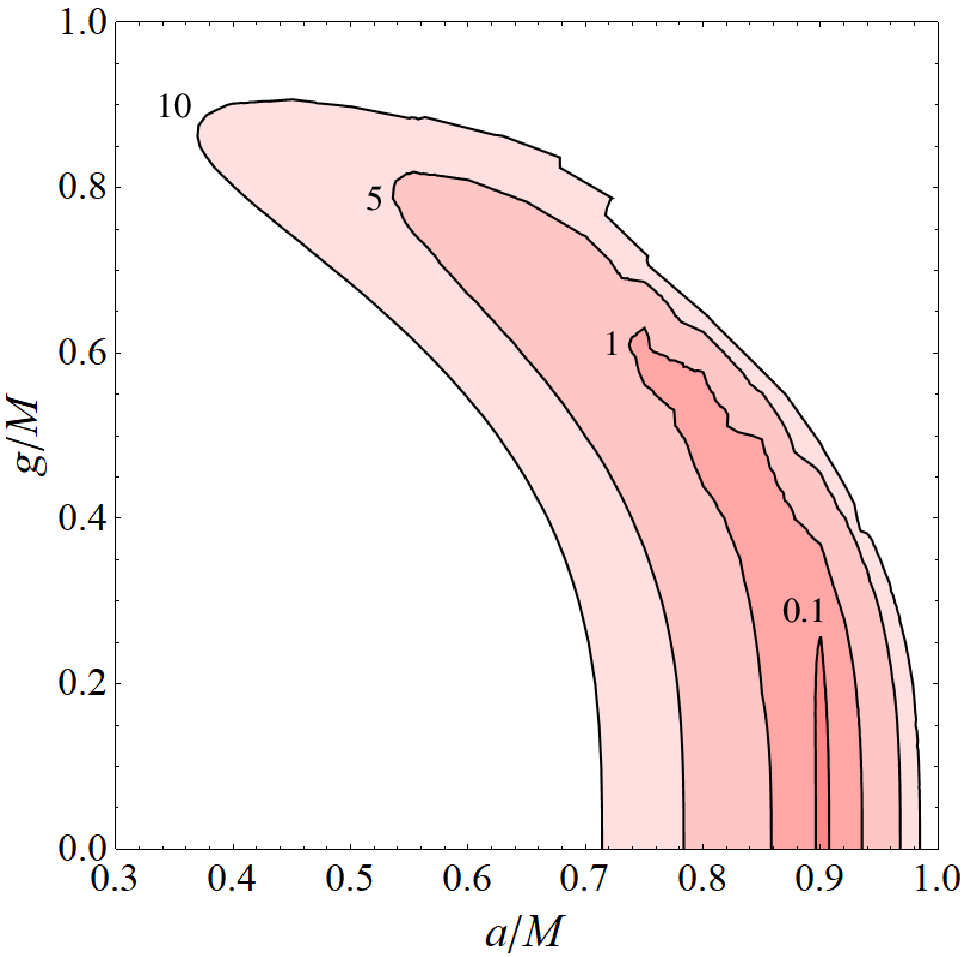}
	\end{tabular}
	\caption{ $\chi ^2_{\text{red}}$ between rotating Hayward black hole and Kerr black hole shadows as a function of ($a,g$) for $a_K=0.20M$, $0.50M$, $0.80M$ and $0.90M$.}   \label{HayFig1}
\end{center}
\end{figure*}
Using Eq.~(\ref{Haywardmass}) in Eq.~(\ref{impactparameter}), we construct the shadow and calculate the corresponding observables $A$ and $D$. To see whether the rotating Hayward black hole (model) shadow can imitate the Kerr black hole (injection) shadow, the $\chi ^2_{\text{red}}$ is calculated between them. The minimized $\chi_{\text{red}}^2$ and the extracted values of the best-fit model spin parameter $a$ as a function of injected spin $a_K$ and for different values of $g$ are shown in Fig.~\ref{HaywardFig}. The $\chi_{\text{red}}^2$ varies nonuniformly with $a_K$, which infers the degree of similarities between the two shadows. Figure \ref{HaywardFig} deduces that the model shadow with the small values of $g$,  can well capture the injection, such that $\chi_{\text{red}}^2\approx 0$ and the extracted values of model spin are close to the injected spin values. Whereas for moderately high values of $g$ ($g\sim 0.50 M$), the model can fit with the injection only for the intermediate values of the injected spin, such that it can not resemble the shadows of very slowly or rapidly rotating injections. Furthermore, for very large values of $g$ ($g\gtrsim 0.70 M$), the two shadows are clearly distinguishable as $\chi_{\text{red}}^2>1$ for all injected spin values ($0\leq a_K\leq M$). The extracted best-fit values of the model spin are biased from the injected spins, and this disparateness is more prominent for higher values of $g$. In addition, constraints are placed on the parameters ($a,g$) for which the model can get fit with the given injection shadow. Figure \ref{HayFig1} shows that for a given shadow injection, there is a finite parameter space for which the model shadow is consistent with the injected shadow. This parameter space decreases with increasing injected spin. Thus, the rapidly rotating Kerr black hole shadows can also be interpreted with the rotating Hayward black hole. 
\subsection{Non-singular black holes}
Bardeen and Hayward regular black holes have an asymptotically de-Sitter core. The next regular model \citep{Ghosh:2014pba,Culetu:2014lca}, which we call non-singular for brevity, is a novel class of regular black hole with an asymptotic Minkowski core \citep{Simpson:2019mud}. While these non-singular models share many features with Bardeen and Hayward black holes, there are also notable differences, especially at the deep core \citep{Simpson:2019mud}. The mass function of rotating non-singular black hole reads  \citep{Ghosh:2014pba}
\begin{eqnarray}
m(r)=Me^{-g^2/2Mr},
\end{eqnarray}
where $g$ is the NED charge. As per the previous section, we calculate the observables $A$ and $D$ for the shadows of rotating non-singular (model) and Kerr (injection) black holes. 
\begin{figure*}
\begin{center}	
	\centering
	\begin{tabular}{c c}
		\includegraphics[scale=0.73]{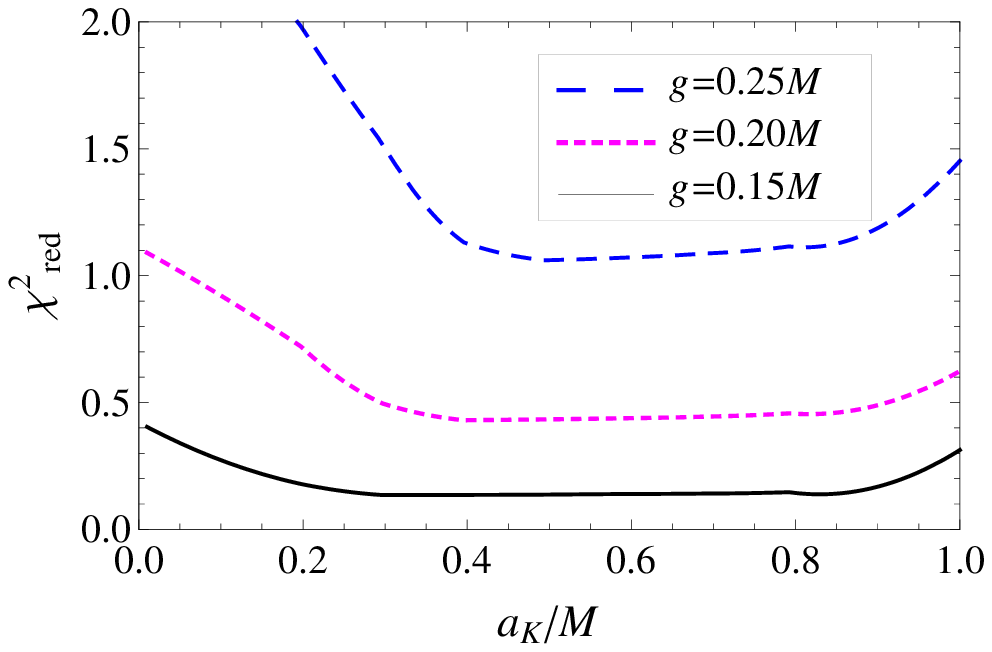} &
		\includegraphics[scale=0.73]{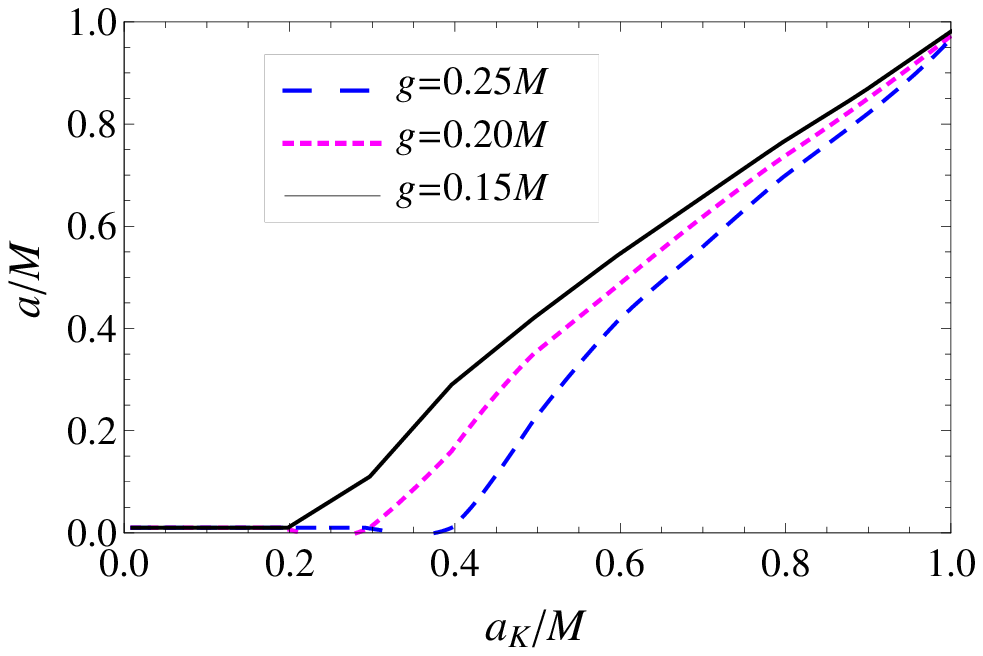}
	\end{tabular}
	\caption{ Minimized $\chi ^2_{\text{red}}$ (left panel) and extracted values of the rotating non-singular black hole spin parameter $a$ (right panel) for the best-fit model shadow as a function of the Kerr spin $a_K$.} \label{NSFig}
	\begin{tabular}{c c c}
		\includegraphics[scale=0.75]{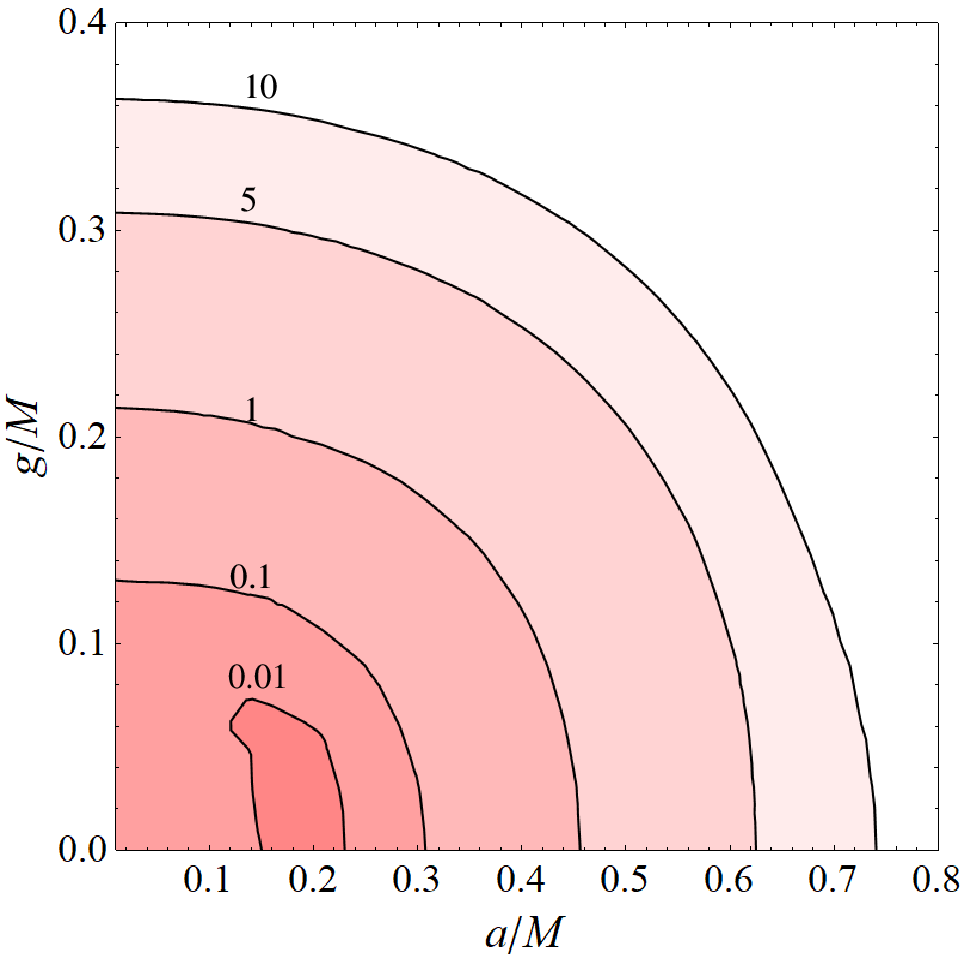} &
		\includegraphics[scale=0.75]{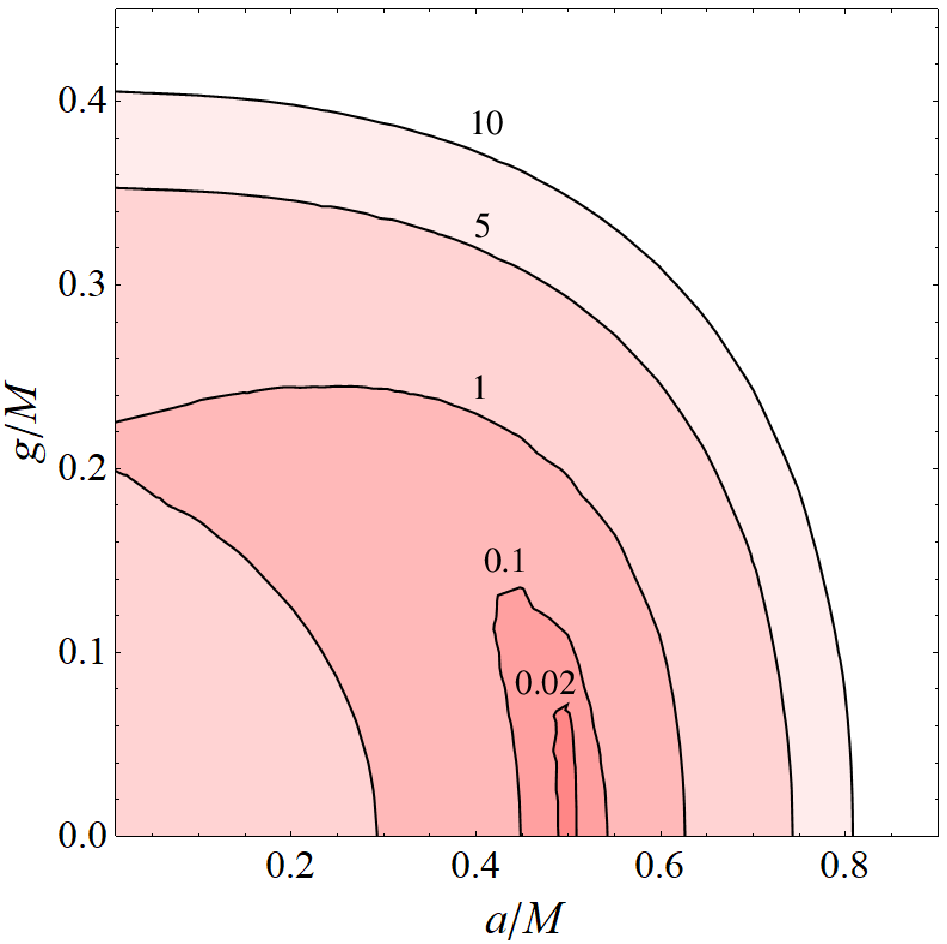}\\
		\includegraphics[scale=0.75]{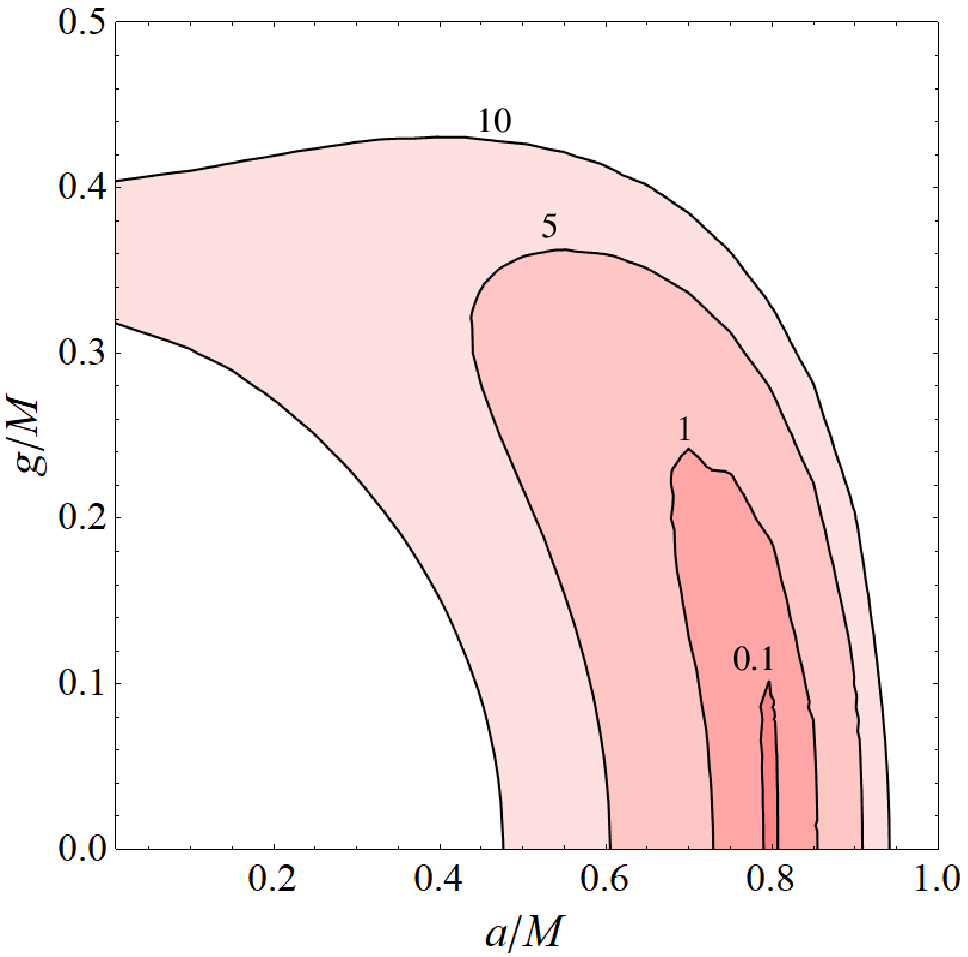}&
		\includegraphics[scale=0.75]{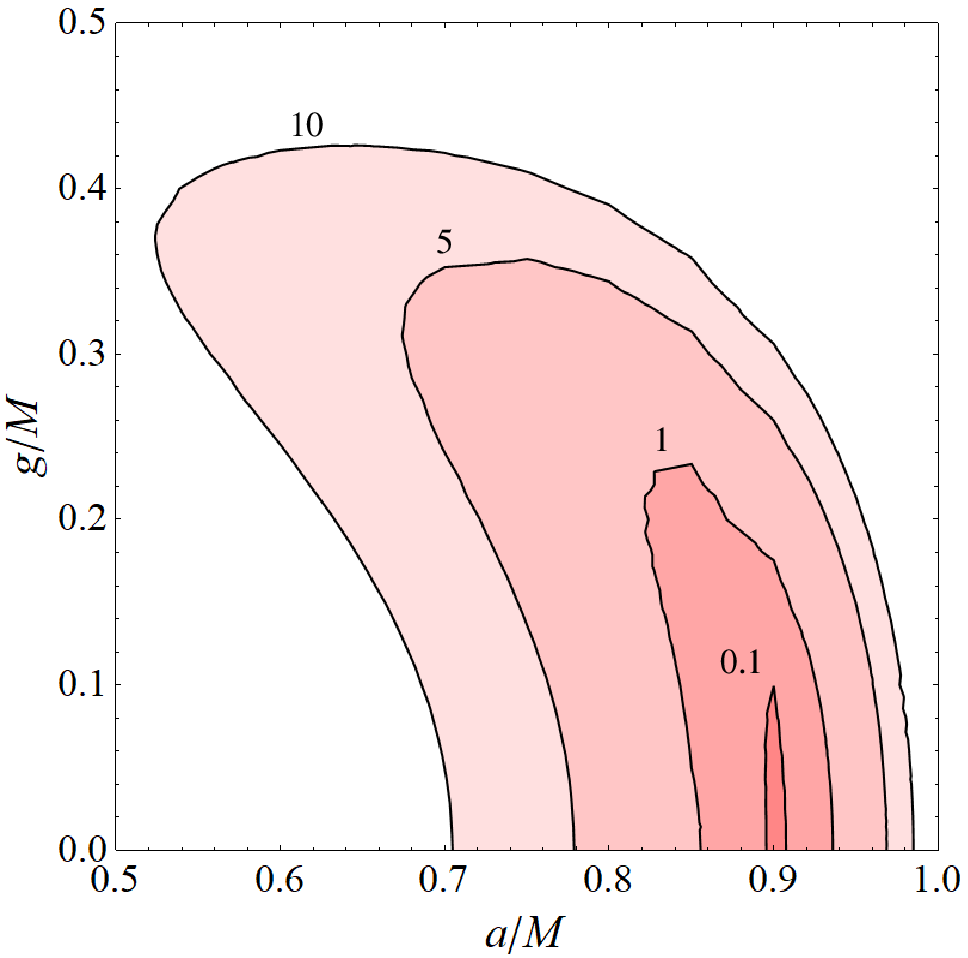}
	\end{tabular}
	\caption{ $\chi ^2_{\text{red}}$ between rotating non-singular black hole and Kerr black hole shadows as a function of ($a,g$) for $a_K=0.20M$, $0.50M$, $0.80M$ and $0.90M$.}   \label{NSfig1}
\end{center}
\end{figure*}
In turn, the $\chi ^2_{\text{red}}$ between the model and injection shadows is minimized to determine the best-fit model spin.
The behavior of  $\chi ^2_{\text{red}}$ with varying $a_K$ is similar to that for rotating Bardeen and Hayward black holes, as the most substantial gain in this model fitting occurs for the intermediate values of $a_K$ (see Fig.~\ref{NSFig}). The minimized $\chi ^2_{\text{red}}$ further increase with $g$. For instance, injected shadow with $a_K=0.6M$, within the current observational uncertainties, is indiscernible from model shadows with spins $a= 0.542M, 0.49M\,\text{and} \,\, 0.417M,$ respectively, for $g=0.15M,0.20M$ and $0.25M$. The best-fit model spins are lower from the injected spin, and the biasness is higher for large $g$. Moreover, the model parameter space ($a,g$), which minimizes the $\chi_{\text{red}}^2$ for $a_K=0.20M,\, 0.50M,\, 0.80M\,\text{and} \,\, 0.90M$, is shown in Fig.~\ref{NSfig1}; this infers that model shadows can resemble the injected shadows, i.e., $\chi^2_{\text{red}}\leq 1$, over the finite parameter space. However, this parameter space ($a,g$) decreases with increasing injected spin $a_K$. Therefore, rapidly rotating Kerr black hole injected shadows placed stringent bounds on rotating non-singular black hole parameters, as the near-extremal values of $a_K$ maximize the general-relativistic effects around the black holes.

\section{Observational constraints from the  M87* black hole Shadow }\label{sect5}
\begin{figure*}
\begin{center}	
\begin{tabular}{c c c}
    \includegraphics[scale=0.5]{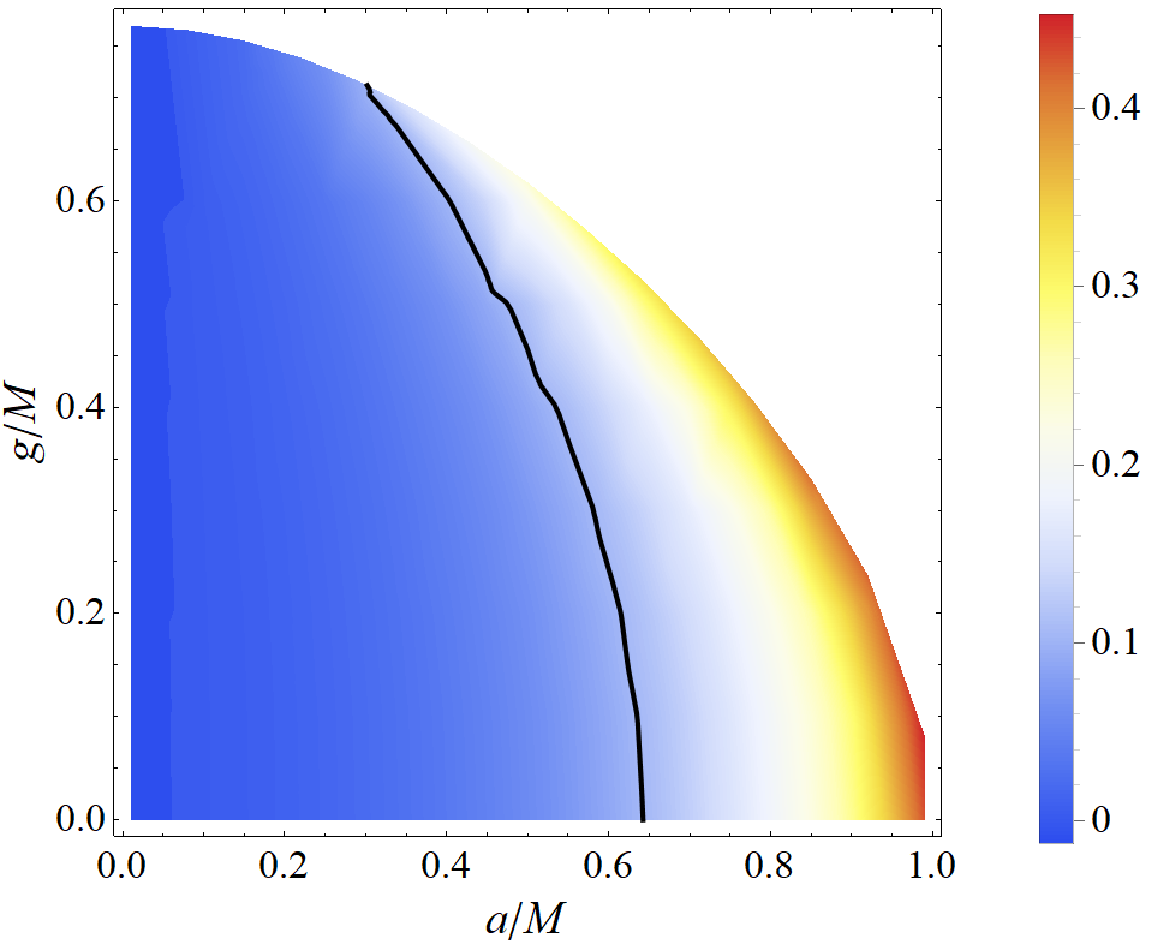}&
	\includegraphics[scale=0.5]{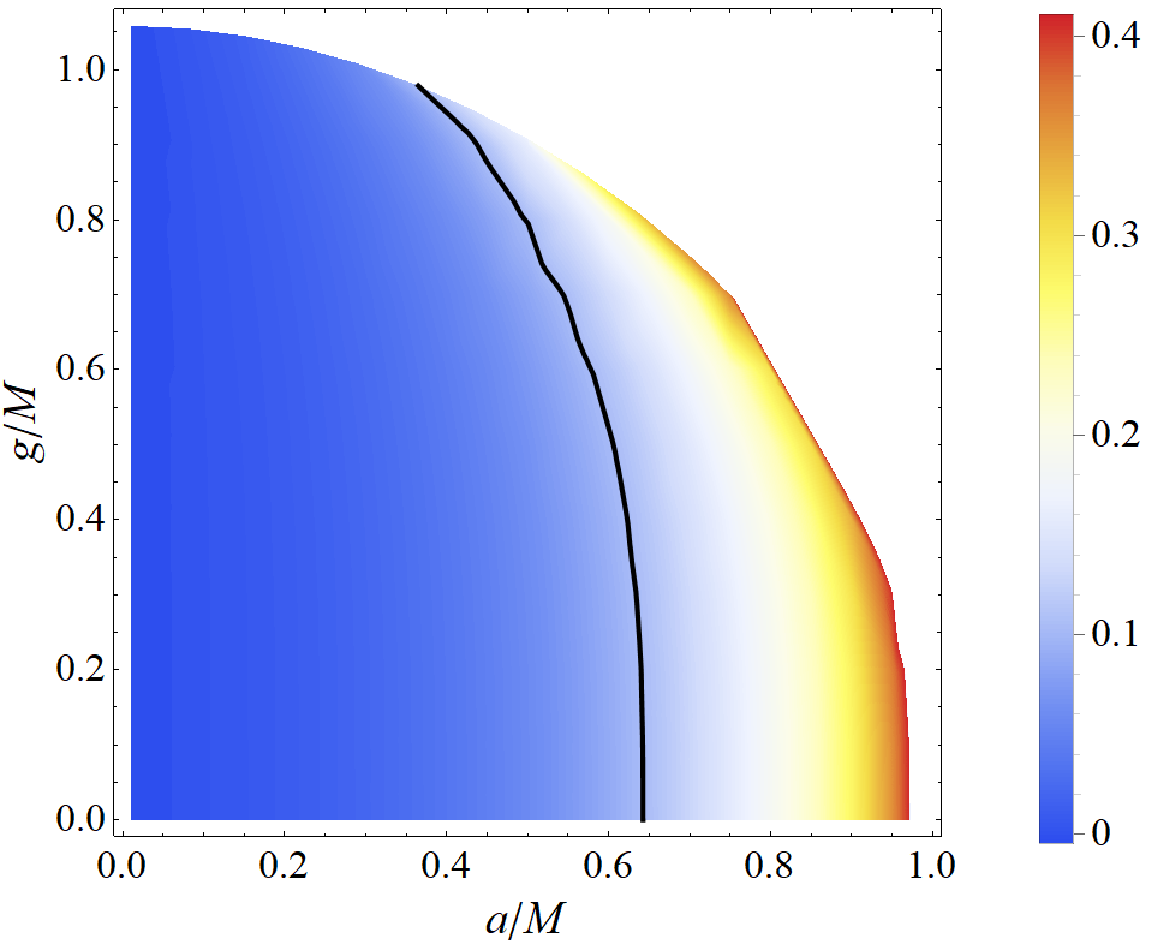}&
	\includegraphics[scale=0.5]{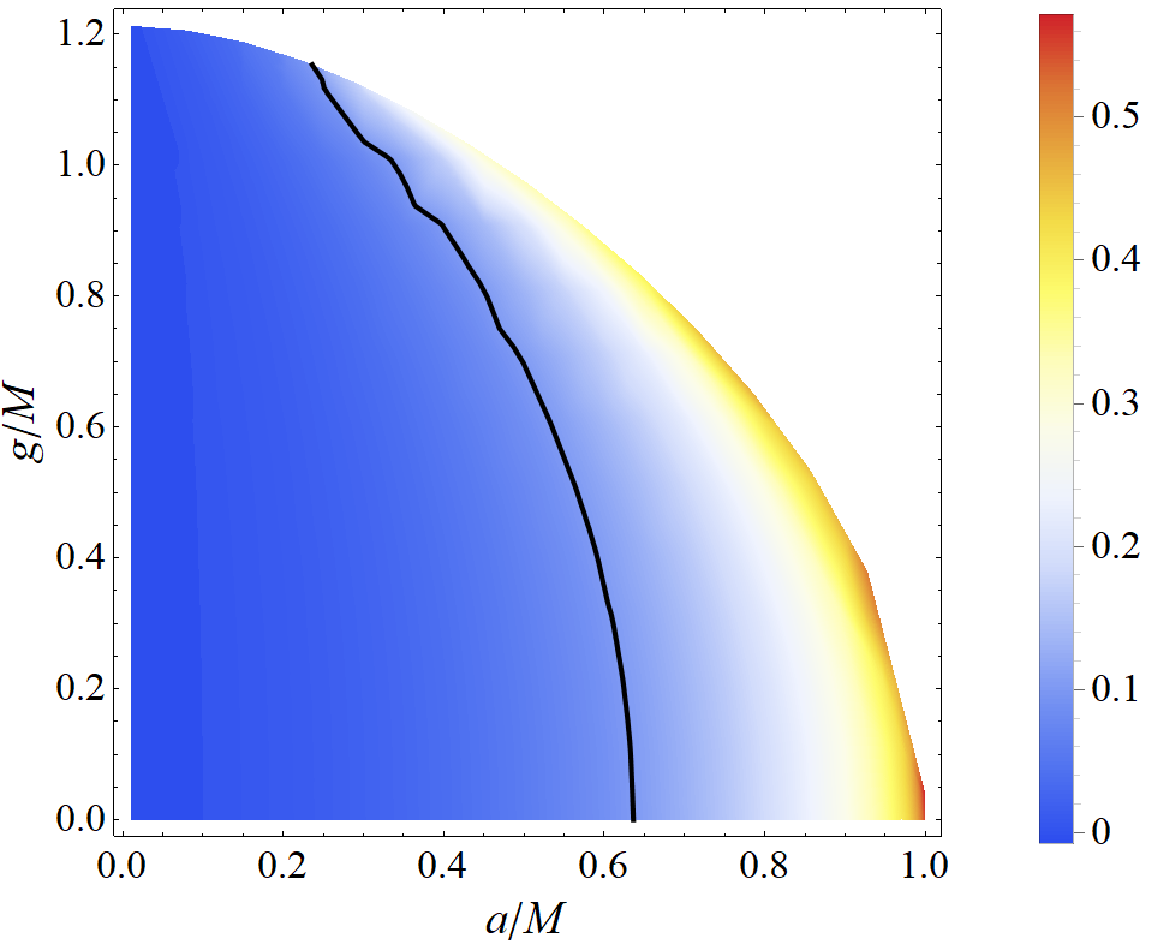}
\end{tabular}
	\caption{Circularity deviation $\Delta C$ as a function of $(a,g)$ for rotating Bardeen (left), rotating Hayward (middle), and rotating non-singular (right) black holes. Black solid line in each plot corresponds to the $\Delta C=0.10$.}\label{M87}
\begin{tabular}{c c c}
	\includegraphics[scale=0.5]{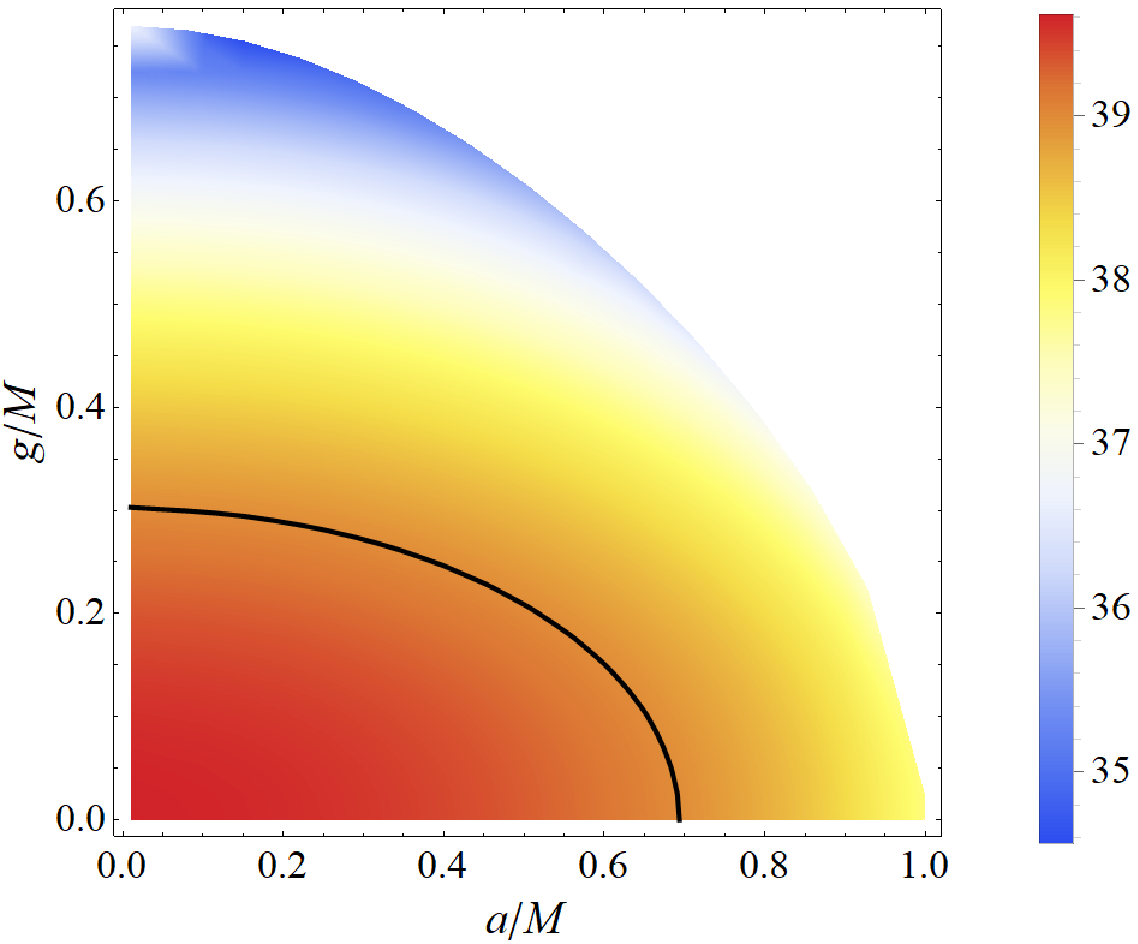}&
	\includegraphics[scale=0.5]{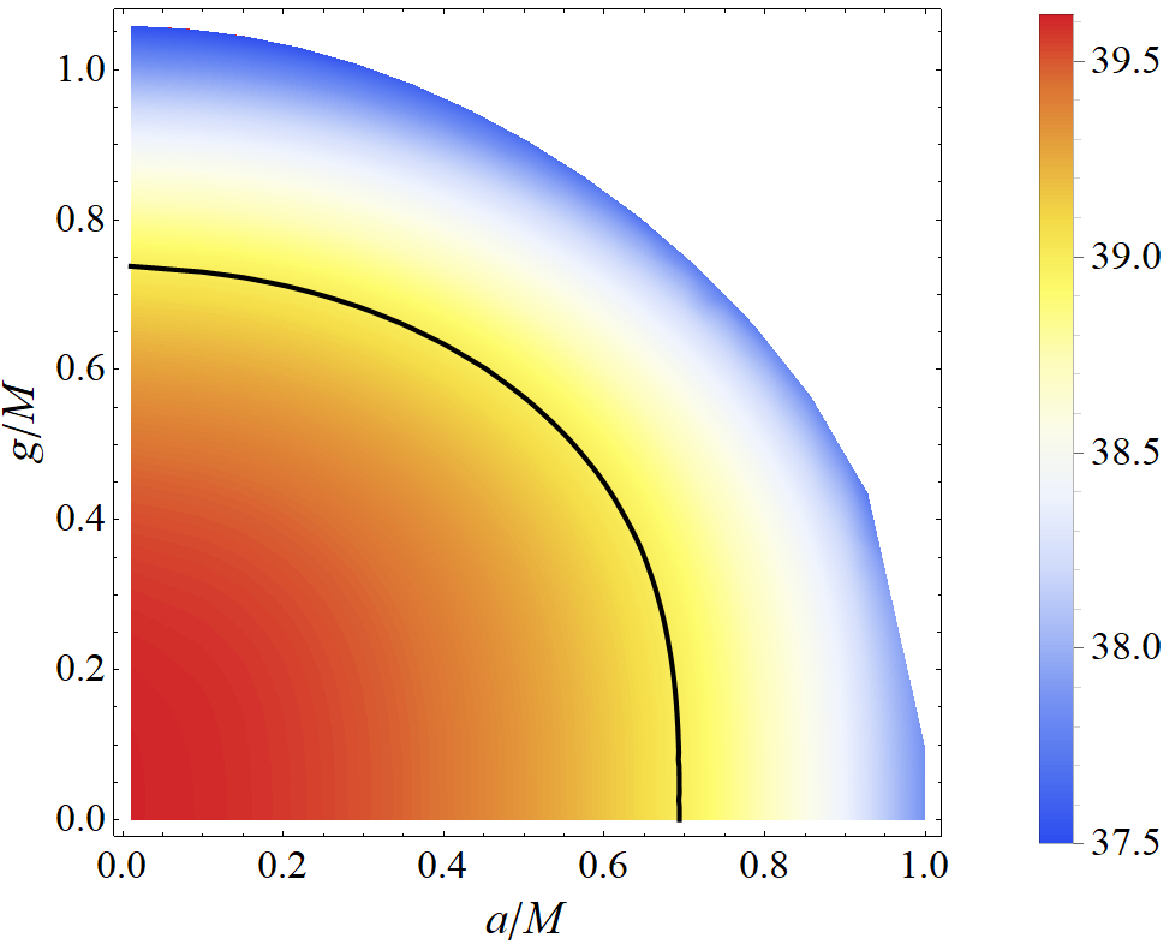}&
	\includegraphics[scale=0.5]{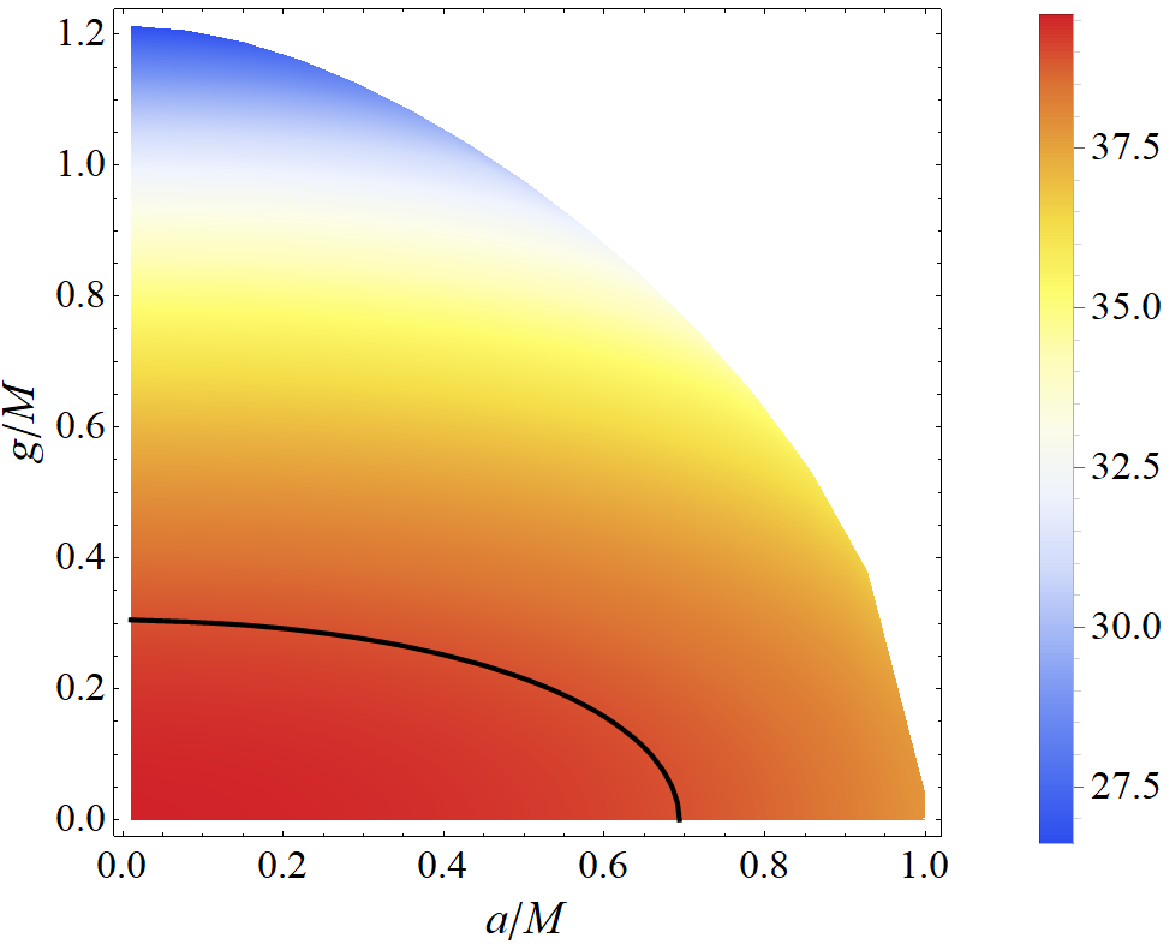}
\end{tabular}
	\caption{Shadow angular diameter $\theta_d$ as a function of $(a,g)$ for rotating Bardeen (left), rotating Hayward (middle), and rotating non-singular (right) black holes. Black solid line in each plot corresponds to the $\theta_d=39\,\mu$as.}\label{M87Ang}
\end{center}
\end{figure*}

Recently, the EHT Collaboration, for the first time, has revealed the 1.3 mm image of the supermassive black hole M87* \citep{Akiyama:2019cqa,Akiyama:2019fyp,Akiyama:2019eap}. Even though the observed image of M87* black hole is consistent with the predicted shadow for the Kerr black hole, because of the uncertainty in the inferred black hole rotation parameter it may be difficult to ignore modified theories black holes \citep{Akiyama:2019cqa,Kumar:2019pjp,Cunha:2019ikd,Vagnozzi:2019apd,Kumar:2019ohr,Neves:2019lio,Allahyari:2019jqz}, e.g., braneworld black holes with negative tidal charge \citep{Banerjee:2019nnj}, and superspinors within finite parameter space \citep{Bambi:2019tjh} can explain the M87* black hole shadow observables. We consider the M87* black hole as the rotating regular black hole and used the shadow observables, namely, angular size and asymmetry, to put constraints on the parameter space ($a,g$) that at best can describe the observed shadow. The shadow is parameterized by ($R(\varphi),\varphi$), where $R(\varphi)$ is the silhouette radial distance from the shadow center ($\alpha_C,\beta_C$), and $\varphi$ is the angular coordinate. The shadow average radius $\bar{R}$ is defined as \citep{Johannsen:2010ru}
\begin{equation}
\bar{R}=\frac{1}{2\pi}\int_{0}^{2\pi} R(\varphi) d\varphi,
\end{equation}
with
\begin{equation}R(\varphi)= \sqrt{(\alpha-\alpha_C)^2+(\beta-\beta_C)^2},\quad \varphi\equiv \tan^{-1}\left(\frac{\beta}{\alpha-\alpha_C}\right)\nonumber,
\end{equation}
and ($\alpha_C,\beta_C$) can be interpreted as the shadow displacement from the black hole center $(0,0)$, such that due to the intrinsic axisymmetry the vertical displacement is zero, i.e., $\beta_C=0$ and $\alpha_C$ is 
\begin{equation}
\alpha_C=\frac{\mid \alpha_{r}+\alpha_{l}\mid}{2}.
\end{equation}
Further, we consider the circularity deviation $\Delta C$ as a measures of the shadow distortion from a perfect circle  \citep{Johannsen:2010ru,Johannsen:2015qca}
\begin{equation}
\Delta C=2\sqrt{\frac{1}{2\pi}\int_0^{2\pi}\left(R(\varphi)-\bar{R}\right)^2d\varphi},
\end{equation}
clearly $\Delta C=0$ for non-rotating black holes that have circular shadows. The EHT data analysis reveals that the circularity deviation in the observed image of the M87* black hole is $\Delta C\leq 0.10$ \citep{Akiyama:2019cqa,Akiyama:2019fyp,Akiyama:2019eap}. The effects of the interplay between spin $a$ and deviation parameter $g$ on the $\Delta C$ for rotating regular black holes are analyzed and depicted in Fig.~\ref{M87}.  
It is clear from Fig.~\ref{M87} that the bound $\Delta C\leq 0.10$ merely constrains the rotating regular black hole parameter space ($a,g$), as all theoretically allowed values of $g$ are permissible for the observed asymmetry of M87*, viz., $g\leq 0.7698M$ for the Bardeen black hole, $g\leq 1.0582M$ for the Hayward black hole, and $g\leq 1.2130M$ for the non-singular black hole. Further, the deduced angular diameter of the M87* black hole shadow can be used to constrain the parameters. 
The shadow areal radius $R_s=\sqrt{A/\pi}$ corresponds to the angular diameter $\theta_d$ in the sky 
\begin{equation}
\theta_d=2\frac{R_s}{d},
\end{equation}    
which depends on the black hole mass $M=(6.5\pm 0.7)\times 10^{9} M_{\odot}$ and its distance from the Earth $d=16.8$ Mpc. The emission ring diameter in the observed M87* black hole shadow is $\theta_d=42\pm 3\, \mu$as, though the emission region is not restricted to lie exactly at the photon ring and preferentially falls outside it. The geometric crescent model accounting for the emission in the general-relativistic-magnetohydrodynamics simulated images and observational uncertainties lead to this offset between the two. The angular diameters of the rotating regular black hole shadows are calculated over the parameter space ($a,g$), and the results are illustrated in Fig.~\ref{M87Ang}.
For the aforementioned black hole mass and distance, the non-rotating Schwarzschild black hole ($a=0,g=0$) cast the largest shadow with angular diameter $\theta_d=39.6192\,\mu$as, which falls within the $1\sigma$ confidence region. The angular diameter $\theta_d$ as a function of ($a,g$) is presented in Fig.~\ref{M87Ang} with the black solid line corresponding to the $\theta_d=39\, \mu$as. As shown in Fig.~\ref{M87Ang}, the angular diameter of the rotating regular black holes shadows over a finite parameter space is remarkably consistent with the observed angular diameter of the M87* black hole, within the $1\sigma$ confidence level. This in turn strongly constrains the parameter $g$, viz., $g\leq 0.30182M$ for the Bardeen black hole, $g\leq 0.73627M$ for the Hayward black hole, and $g\leq 0.30461M$ for the non-singular black hole. The rotating regular black holes with a very large value of $g$ can not account for the observed angular size of M87*. Thus, from Figs.~\ref{M87} and \ref{M87Ang}, it is evident that the M87* shadow shape and size could be explained with the rotating regular black holes, and they can be strong candidates for the astrophysical black holes.
\section{Conclusion}\label{sect6}
In this paper, we present the systematic bias analysis for
testing the rotating regular black hole (non-Kerr) metrics to analyze the deviation of their shadows from that of the Kerr black hole, which is detectable with the EHT observations.
The shape and size of the black hole shadow can be very well determined only by the spacetime geometry, the observer's viewing angle, and by identifying the photons captured region, which ascertains that the shadow can be regarded as a potential tool for this. The size and shape of the black hole shadow are characterized by two observables, viz., area $A$ and oblateness $D$ \citep{Kumar:2018ple}. It turns out that the same shadow observable ($A, D$) can be associated with several black hole models with significantly different parameters \citep{Kumar:2018ple}. In particular, we have investigated whether the black hole shadow, using these observables, can determine if astrophysical black holes are indeed non-Kerr focusing on well-motivated three rotating regular black holes having an additional deviation parameter $g$ due to the NED charge. We analyzed shadows produced by rotating regular black holes and compared them with those for the Kerr black holes using systematic bias analysis. We showed that rotating regular black holes, depending on the values of $g$, in some cases, cause shadows that are very similar to those produced by the Kerr black holes ($\chi^2_{\text{red}} < 1$), but in other cases, the two would be clearly distinguishable ($\chi^2_{\text{red}}> 1$). To illustrate on the above point, model shadows, viz., Bardeen black holes  ($g\lesssim 0.26 M$), Hayward black holes  ($g\lesssim 0.65 M$), and  non-singular black holes ($g\lesssim 0.25 M$), can well capture the Kerr black hole shadows within the current observational uncertainties (see Figs.~\ref{BardeenFig1}, \ref{HayFig1} and \ref{NSfig1} ). The spin $a$ of the best-fit models, which minimized the  $\chi^2_{\text{red}}$, are found to be biased from the injected spin values $a_K$, and that becomes more biased with increasing $g$.  In particular, model shadows with higher $g$ resembled the injected shadows only for the intermediate  $a_K$; therefore the rotating regular black holes cannot mimic the slowly or rapidly rotating Kerr black hole shadows. Whereas for sufficiently large values of $g$, model shadows significantly differed from the injected shadows, and the current observational facilities can unambiguously discern  ($\chi^2_{\text{red}}> 1$) the model with injection shadows.  

In turn, the M87* black hole shadow observables $(\Delta C, \theta_d)$ are found  to be consistent with the rotating regular black hole shadows within the finite particular parameter space ($a,g$) (see Fig.~\ref{M87} and \ref{M87Ang}). Further, according to our analysis, Bardeen black holes ($g\leq 0.30182M$), Hayward black holes ($g\leq 0.73627M$), and non-singular black holes ($g\leq 0.30461M$), within the $1\sigma$ region for $\theta_d= 39\, \mu$as,  are consistent with the observed angular diameter of M87* black hole implying that the rotating regular black holes can be strong viable candidates for the astrophysical black holes.

It is straightforward to employ the formalism to any other stationary, axisymmetric, and asymptotically flat black hole metric. In particular, it will be physically much more interesting to consider other non-Kerr black hole metrics, arising in modified gravities such as that proposed by Johannsen and Psaltis (\citeyear{Johannsen:2010ru}), to compare with shadows produced by Kerr black holes for getting more insights. 

\section{Acknowledgment}
S.G.G. would like to thank DST INDO-SA bilateral project DST/INT/South Africa/P-06/2016, SERB-DST for the ASEAN project IMRC/AISTDF/CRD/2018/000042  and also to IUCAA, Pune for the hospitality while this work was being done. R.K. would like to thank UGC for providing SRF.




\bibliography{ChiSqr}{}

\begin{thebibliography}{}
\expandafter\ifx\csname natexlab\endcsname\relax\def\natexlab#1{#1}\fi
\providecommand{\url}[1]{\href{#1}{#1}}
\providecommand{\dodoi}[1]{doi:~\href{http://doi.org/#1}{\nolinkurl{#1}}}
\providecommand{\doeprint}[1]{\href{http://ascl.net/#1}{\nolinkurl{http://ascl.net/#1}}}
\providecommand{\doarXiv}[1]{\href{https://arxiv.org/abs/#1}{\nolinkurl{https://arxiv.org/abs/#1}}}

\bibitem[{Abdujabbarov {et~al.}(2016)Abdujabbarov, Amir, Ahmedov, \&
  Ghosh}]{Abdujabbarov:2016hnw}
Abdujabbarov, A., Amir, M., Ahmedov, B., \& Ghosh, S.~G. 2016, Phys. Rev., D93,
  104004, \dodoi{10.1103/PhysRevD.93.104004}

\bibitem[{Abdujabbarov {et~al.}(2015)Abdujabbarov, Atamurotov, Dadhich,
  Ahmedov, \& Stuchlik}]{Abdujabbarov:2015rqa}
Abdujabbarov, A., Atamurotov, F., Dadhich, N., Ahmedov, B., \& Stuchlik, Z.
  2015, Eur. Phys. J., C75, 399, \dodoi{10.1140/epjc/s10052-015-3604-5}

\bibitem[{Akiyama {et~al.}(2019{\natexlab{a}})}]{Akiyama:2019cqa}
Akiyama, K., {et~al.} 2019{\natexlab{a}}, Astrophys. J., 875, L1,
  \dodoi{10.3847/2041-8213/ab0ec7}

\bibitem[{Akiyama {et~al.}(2019{\natexlab{b}})}]{Akiyama:2019fyp}
---. 2019{\natexlab{b}}, Astrophys. J., 875, L5,
  \dodoi{10.3847/2041-8213/ab0f43}

\bibitem[{Akiyama {et~al.}(2019{\natexlab{c}})}]{Akiyama:2019eap}
---. 2019{\natexlab{c}}, Astrophys. J., 875, L6,
  \dodoi{10.3847/2041-8213/ab1141}

\bibitem[{Allahyari {et~al.}(2020)Allahyari, Khodadi, Vagnozzi, \&
  Mota}]{Allahyari:2019jqz}
Allahyari, A., Khodadi, M., Vagnozzi, S., \& Mota, D.~F. 2020, JCAP, 02, 003,
  \dodoi{10.1088/1475-7516/2020/02/003}

\bibitem[{Amarilla \& Eiroa(2012)}]{Amarilla:2011fx}
Amarilla, L., \& Eiroa, E.~F. 2012, Phys. Rev., D85, 064019,
  \dodoi{10.1103/PhysRevD.85.064019}

\bibitem[{Amarilla {et~al.}(2010)Amarilla, Eiroa, \& Giribet}]{Amarilla:2010zq}
Amarilla, L., Eiroa, E.~F., \& Giribet, G. 2010, Phys. Rev., D81, 124045,
  \dodoi{10.1103/PhysRevD.81.124045}

\bibitem[{Amir {et~al.}(2016)Amir, Ahmed, \& Ghosh}]{Amir:2016nti}
Amir, M., Ahmed, F., \& Ghosh, S.~G. 2016, Eur. Phys. J., C76, 532,
  \dodoi{10.1140/epjc/s10052-016-4365-5}

\bibitem[{Amir \& Ghosh(2015)}]{Amir:2015pja}
Amir, M., \& Ghosh, S.~G. 2015, JHEP, 07, 015, \dodoi{10.1007/JHEP07(2015)015}

\bibitem[{Amir \& Ghosh(2016)}]{Amir:2016cen}
---. 2016, Phys. Rev., D94, 024054, \dodoi{10.1103/PhysRevD.94.024054}

\bibitem[{Amir {et~al.}(2018)Amir, Singh, \& Ghosh}]{Amir:2017slq}
Amir, M., Singh, B.~P., \& Ghosh, S.~G. 2018, Eur. Phys. J., C78, 399,
  \dodoi{10.1140/epjc/s10052-018-5872-3}

\bibitem[{Atamurotov {et~al.}(2013)Atamurotov, Abdujabbarov, \&
  Ahmedov}]{Atamurotov:2013sca}
Atamurotov, F., Abdujabbarov, A., \& Ahmedov, B. 2013, Phys. Rev., D88, 064004,
  \dodoi{10.1103/PhysRevD.88.064004}

\bibitem[{Ayon-Beato \& Garcia(1998)}]{AyonBeato:1998ub}
Ayon-Beato, E., \& Garcia, A. 1998, Phys. Rev. Lett., 80, 5056,
  \dodoi{10.1103/PhysRevLett.80.5056}

\bibitem[{Ayon-Beato \& Garcia(1999)}]{AyonBeato:1999ec}
---. 1999, Gen. Rel. Grav., 31, 629, \dodoi{10.1023/A:1026640911319}

\bibitem[{Ayon-Beato \& Garcia(2000)}]{AyonBeato:2000zs}
---. 2000, Phys. Lett., B493, 149, \dodoi{10.1016/S0370-2693(00)01125-4}

\bibitem[{Ayzenberg \& Yunes(2018)}]{Ayzenberg:2018jip}
Ayzenberg, D., \& Yunes, N. 2018, Class. Quant. Grav., 35, 235002,
  \dodoi{10.1088/1361-6382/aae87b}

\bibitem[{Azreg-Ainou(2014)}]{Azreg-Ainou:2014pra}
Azreg-Ainou, M. 2014, Phys. Rev., D90, 064041,
  \dodoi{10.1103/PhysRevD.90.064041}

\bibitem[{Bambi(2014)}]{Bambi:2014nta}
Bambi, C. 2014, Phys. Lett., B730, 59, \dodoi{10.1016/j.physletb.2014.01.037}

\bibitem[{Bambi \& Barausse(2011)}]{Bambi:2011jq}
Bambi, C., \& Barausse, E. 2011, Astrophys. J., 731, 121,
  \dodoi{10.1088/0004-637X/731/2/121}

\bibitem[{Bambi \& Freese(2009)}]{Bambi:2008jg}
Bambi, C., \& Freese, K. 2009, Phys. Rev., D79, 043002,
  \dodoi{10.1103/PhysRevD.79.043002}

\bibitem[{Bambi {et~al.}(2019)Bambi, Freese, Vagnozzi, \&
  Visinelli}]{Bambi:2019tjh}
Bambi, C., Freese, K., Vagnozzi, S., \& Visinelli, L. 2019, Phys. Rev., D100,
  044057, \dodoi{10.1103/PhysRevD.100.044057}

\bibitem[{Bambi \& Modesto(2013)}]{Bambi:2013ufa}
Bambi, C., \& Modesto, L. 2013, Phys. Lett., B721, 329,
  \dodoi{10.1016/j.physletb.2013.03.025}

\bibitem[{Banerjee {et~al.}(2020)Banerjee, Chakraborty, \&
  SenGupta}]{Banerjee:2019nnj}
Banerjee, I., Chakraborty, S., \& SenGupta, S. 2020, Phys. Rev. D, 101, 041301,
  \dodoi{10.1103/PhysRevD.101.041301}

\bibitem[{Bardeen(1968)}]{bardeen1968non}
Bardeen, J. 1968, Tbilisi, USSR

\bibitem[{Bardeen(1973)}]{bardeen1973}
---. 1973, Black Holes, edited by C. DeWitt and BS DeWitt,  Gordon and Breach,
  New York

\bibitem[{Beckwith \& Done(2005)}]{Beckwith:2004ae}
Beckwith, K., \& Done, C. 2005, Mon. Not. Roy. Astron. Soc., 359, 1217,
  \dodoi{10.1111/j.1365-2966.2005.08980.x}

\bibitem[{Berej {et~al.}(2006)Berej, Matyjasek, Tryniecki, \&
  Woronowicz}]{Berej:2006cc}
Berej, W., Matyjasek, J., Tryniecki, D., \& Woronowicz, M. 2006, Gen. Rel.
  Grav., 38, 885, \dodoi{10.1007/s10714-006-0270-9}

\bibitem[{Bronnikov(2001)}]{Bronnikov:2000vy}
Bronnikov, K.~A. 2001, Phys. Rev., D63, 044005,
  \dodoi{10.1103/PhysRevD.63.044005}

\bibitem[{Bronnikov(2017)}]{Bronnikov:2017tnz}
---. 2017, Phys. Rev., D96, 128501, \dodoi{10.1103/PhysRevD.96.128501}

\bibitem[{Bronnikov \& Fabris(2006)}]{Bronnikov:2005gm}
Bronnikov, K.~A., \& Fabris, J.~C. 2006, Phys. Rev. Lett., 96, 251101,
  \dodoi{10.1103/PhysRevLett.96.251101}

\bibitem[{Burinskii \& Hildebrandt(2002)}]{Burinskii:2002pz}
Burinskii, A., \& Hildebrandt, S.~R. 2002, Phys. Rev., D65, 104017,
  \dodoi{10.1103/PhysRevD.65.104017}

\bibitem[{Cardoso \& Pani(2019)}]{Cardoso:2019rvt}
Cardoso, V., \& Pani, P. 2019, Living Rev. Rel., 22, 4,
  \dodoi{10.1007/s41114-019-0020-4}

\bibitem[{Carter(1968)}]{Carter:1968rr}
Carter, B. 1968, Phys. Rev., 174, 1559, \dodoi{10.1103/PhysRev.174.1559}

\bibitem[{Carter(1971)}]{Carter:1971zc}
---. 1971, Phys. Rev. Lett., 26, 331, \dodoi{10.1103/PhysRevLett.26.331}

\bibitem[{Chandrasekhar(1985)}]{Chandrasekhar:1985kt}
Chandrasekhar, S. 1985, {The mathematical theory of black holes} (Oxford:
  Oxford Univ. Press)

\bibitem[{Culetu(2015)}]{Culetu:2014lca}
Culetu, H. 2015, Int. J. Theor. Phys., 54, 2855,
  \dodoi{10.1007/s10773-015-2521-6}

\bibitem[{Cunha {et~al.}(2017)Cunha, Herdeiro, Kleihaus, Kunz, \&
  Radu}]{Cunha:2016wzk}
Cunha, P. V.~P., Herdeiro, C. A.~R., Kleihaus, B., Kunz, J., \& Radu, E. 2017,
  Phys. Lett., B768, 373, \dodoi{10.1016/j.physletb.2017.03.020}

\bibitem[{Cunha {et~al.}(2019)Cunha, Herdeiro, \& Radu}]{Cunha:2019ikd}
Cunha, P. V.~P., Herdeiro, C. A.~R., \& Radu, E. 2019, Universe, 5, 220,
  \dodoi{10.3390/universe5120220}

\bibitem[{{Cunningham} \& {Bardeen}(1973)}]{CT}
{Cunningham}, C.~T., \& {Bardeen}, J.~M. 1973, \apj, 183, 237,
  \dodoi{10.1086/152223}

\bibitem[{Dymnikova(1992)}]{Dymnikova:1992ux}
Dymnikova, I. 1992, Gen. Rel. Grav., 24, 235, \dodoi{10.1007/BF00760226}

\bibitem[{Dymnikova(2004)}]{Dymnikova:2004zc}
---. 2004, Class. Quant. Grav., 21, 4417, \dodoi{10.1088/0264-9381/21/18/009}

\bibitem[{Eiroa \& Sendra(2013)}]{Eiroa:2013nra}
Eiroa, E.~F., \& Sendra, C.~M. 2013, Phys. Rev., D88, 103007,
  \dodoi{10.1103/PhysRevD.88.103007}

\bibitem[{Falcke \& Markoff(2013)}]{Falcke:2013ola}
Falcke, H., \& Markoff, S.~B. 2013, Class. Quant. Grav., 30, 244003,
  \dodoi{10.1088/0264-9381/30/24/244003}

\bibitem[{Fan \& Wang(2016)}]{Fan:2016hvf}
Fan, Z.-Y., \& Wang, X. 2016, Phys. Rev., D94, 124027,
  \dodoi{10.1103/PhysRevD.94.124027}

\bibitem[{Garcia {et~al.}(2015)Garcia, Hackmann, Kunz, Lämmerzahl, \&
  Macias}]{Garcia:2013zud}
Garcia, A., Hackmann, E., Kunz, J., Lämmerzahl, C., \& Macias, A. 2015, J.
  Math. Phys., 56, 032501, \dodoi{10.1063/1.4913882}

\bibitem[{Ghasemi-Nodehi {et~al.}(2015)Ghasemi-Nodehi, Li, \&
  Bambi}]{Ghasemi-Nodehi:2015raa}
Ghasemi-Nodehi, M., Li, Z., \& Bambi, C. 2015, Eur. Phys. J., C75, 315,
  \dodoi{10.1140/epjc/s10052-015-3539-x}

\bibitem[{Ghosh(2015)}]{Ghosh:2014pba}
Ghosh, S.~G. 2015, Eur. Phys. J., C75, 532,
  \dodoi{10.1140/epjc/s10052-015-3740-y}

\bibitem[{Ghosh \& Maharaj(2015)}]{Ghosh:2014hea}
Ghosh, S.~G., \& Maharaj, S.~D. 2015, Eur. Phys. J., C75, 7,
  \dodoi{10.1140/epjc/s10052-014-3222-7}

\bibitem[{{Gliner}(1966)}]{Gliner}
{Gliner}, E.~B. 1966, Soviet Journal of Experimental and Theoretical Physics,
  22, 378

\bibitem[{Gou {et~al.}(2011)Gou, McClintock, Reid, Orosz, Steiner, Narayan,
  Xiang, Remillard, Arnaud, \& Davis}]{Gou:2011nq}
Gou, L., McClintock, J.~E., Reid, M.~J., {et~al.} 2011, Astrophys. J., 742, 85,
  \dodoi{10.1088/0004-637X/742/2/85}

\bibitem[{Gou {et~al.}(2014)Gou, McClintock, Remillard, Steiner, Reid, Orosz,
  Narayan, Hanke, \& García}]{Gou:2013dna}
Gou, L., McClintock, J.~E., Remillard, R.~A., {et~al.} 2014, Astrophys. J.,
  790, 29, \dodoi{10.1088/0004-637X/790/1/29}

\bibitem[{Grenzebach {et~al.}(2014)Grenzebach, Perlick, \&
  Lämmerzahl}]{Grenzebach:2014fha}
Grenzebach, A., Perlick, V., \& Lämmerzahl, C. 2014, Phys. Rev., D89, 124004,
  \dodoi{10.1103/PhysRevD.89.124004}

\bibitem[{Hawking(1972)}]{Hawking:1971vc}
Hawking, S.~W. 1972, Commun. Math. Phys., 25, 152, \dodoi{10.1007/BF01877517}

\bibitem[{Hayward(2006)}]{Hayward:2005gi}
Hayward, S.~A. 2006, Phys. Rev. Lett., 96, 031103,
  \dodoi{10.1103/PhysRevLett.96.031103}

\bibitem[{Hioki \& Maeda(2009)}]{Hioki:2009na}
Hioki, K., \& Maeda, K.-i. 2009, Phys. Rev., D80, 024042,
  \dodoi{10.1103/PhysRevD.80.024042}

\bibitem[{Israel(1967)}]{Israel:1967wq}
Israel, W. 1967, Phys. Rev., 164, 1776, \dodoi{10.1103/PhysRev.164.1776}

\bibitem[{Israel(1968)}]{Israel:1967za}
---. 1968, Commun. Math. Phys., 8, 245, \dodoi{10.1007/BF01645859}

\bibitem[{Johannsen(2013{\natexlab{a}})}]{Johannsen:2013rqa}
Johannsen, T. 2013{\natexlab{a}}, Phys. Rev., D87, 124017,
  \dodoi{10.1103/PhysRevD.87.124017}

\bibitem[{Johannsen(2013{\natexlab{b}})}]{Johannsen:2015qca}
---. 2013{\natexlab{b}}, Astrophys. J., 777, 170,
  \dodoi{10.1088/0004-637X/777/2/170}

\bibitem[{Johannsen(2016)}]{Johannsen:2016uoh}
---. 2016, Class. Quant. Grav., 33, 124001,
  \dodoi{10.1088/0264-9381/33/12/124001}

\bibitem[{Johannsen \& Psaltis(2010)}]{Johannsen:2010ru}
Johannsen, T., \& Psaltis, D. 2010, Astrophys. J., 718, 446,
  \dodoi{10.1088/0004-637X/718/1/446}

\bibitem[{Junior {et~al.}(2015)Junior, Rodrigues, \& Houndjo}]{Junior:2015fya}
Junior, E. L.~B., Rodrigues, M.~E., \& Houndjo, M. J.~S. 2015, JCAP, 1510, 060,
  \dodoi{10.1088/1475-7516/2015/10/060}

\bibitem[{Jusufi {et~al.}(2018)Jusufi, Ovgun, Saavedra, Vasquez, \&
  Gonzalez}]{Jusufi:2018jof}
Jusufi, K., Ovgun, A., Saavedra, J., Vasquez, Y., \& Gonzalez, P.~A. 2018,
  Phys. Rev., D97, 124024, \dodoi{10.1103/PhysRevD.97.124024}

\bibitem[{Kerr(1963)}]{Kerr:1963ud}
Kerr, R.~P. 1963, Phys. Rev. Lett., 11, 237, \dodoi{10.1103/PhysRevLett.11.237}

\bibitem[{Konoplya \& Zhidenko(2019)}]{Konoplya:2019goy}
Konoplya, R.~A., \& Zhidenko, A. 2019, Phys. Rev., D100, 044015,
  \dodoi{10.1103/PhysRevD.100.044015}

\bibitem[{Kumar \& Ghosh(2020)}]{Kumar:2018ple}
Kumar, R., \& Ghosh, S.~G. 2020, Astrophys. J., 892, 78,
  \dodoi{10.3847/1538-4357/ab77b0}

\bibitem[{Kumar {et~al.}(2019{\natexlab{a}})Kumar, Ghosh, \&
  Wang}]{Kumar:2019pjp}
Kumar, R., Ghosh, S.~G., \& Wang, A. 2019{\natexlab{a}}, Phys. Rev., D100,
  124024, \dodoi{10.1103/PhysRevD.100.124024}

\bibitem[{Kumar {et~al.}(2019{\natexlab{b}})Kumar, Singh, \&
  Ghosh}]{Kumar:2019ohr}
Kumar, R., Singh, B.~P., \& Ghosh, S.~G. 2019{\natexlab{b}}, arXiv, 1904.07652.
\newblock \doarXiv{1904.07652}

\bibitem[{Lamy {et~al.}(2018)Lamy, Gourgoulhon, Paumard, \&
  Vincent}]{Lamy:2018zvj}
Lamy, F., Gourgoulhon, E., Paumard, T., \& Vincent, F.~H. 2018, Class. Quant.
  Grav., 35, 115009, \dodoi{10.1088/1361-6382/aabd97}

\bibitem[{Li \& Bambi(2014)}]{Li:2013jra}
Li, Z., \& Bambi, C. 2014, JCAP, 1401, 041,
  \dodoi{10.1088/1475-7516/2014/01/041}

\bibitem[{Liu {et~al.}(2019)Liu, Ding, \& Jing}]{Liu:2017ifc}
Liu, C., Ding, C., \& Jing, J. 2019, Sci. China Phys. Mech. Astron., 62, 10411,
  \dodoi{10.1007/s11433-018-9243-x}

\bibitem[{Long {et~al.}(2019)Long, Wang, Chen, \& Jing}]{Long:2019nox}
Long, F., Wang, J., Chen, S., \& Jing, J. 2019, JHEP, 10, 269,
  \dodoi{10.1007/JHEP10(2019)269}

\bibitem[{Mishra {et~al.}(2019)Mishra, Chakraborty, \& Sarkar}]{Mishra:2019trb}
Mishra, A.~K., Chakraborty, S., \& Sarkar, S. 2019, Phys. Rev., D99, 104080,
  \dodoi{10.1103/PhysRevD.99.104080}

\bibitem[{Neves(2020)}]{Neves:2019lio}
Neves, J.~C. 2020, Eur. Phys. J. C, 80, 343,
  \dodoi{10.1140/epjc/s10052-020-7913-y}

\bibitem[{Newman {et~al.}(1965)Newman, Couch, Chinnapared, Exton, Prakash, \&
  Torrence}]{Newman:1965my}
Newman, E.~T., Couch, R., Chinnapared, K., {et~al.} 1965, J. Math. Phys., 6,
  918, \dodoi{10.1063/1.1704351}

\bibitem[{Ovgun(2019)}]{Ovgun:2019wej}
Ovgun, A. 2019, Phys. Rev., D99, 104075, \dodoi{10.1103/PhysRevD.99.104075}

\bibitem[{Perlick {et~al.}(2018)Perlick, Tsupko, \&
  Bisnovatyi-Kogan}]{Perlick:2018iye}
Perlick, V., Tsupko, O.~{\relax Yu}., \& Bisnovatyi-Kogan, G.~S. 2018, Phys.
  Rev., D97, 104062, \dodoi{10.1103/PhysRevD.97.104062}

\bibitem[{Robinson(1975)}]{Robinson:1975bv}
Robinson, D.~C. 1975, Phys. Rev. Lett., 34, 905,
  \dodoi{10.1103/PhysRevLett.34.905}

\bibitem[{Rodrigues \& Junior(2017)}]{Rodrigues:2017tfm}
Rodrigues, M.~E., \& Junior, E. L.~B. 2017, Phys. Rev., D96, 128502,
  \dodoi{10.1103/PhysRevD.96.128502}

\bibitem[{Ryan(1995)}]{Ryan:1995wh}
Ryan, F.~D. 1995, Phys. Rev., D52, 5707, \dodoi{10.1103/PhysRevD.52.5707}

\bibitem[{Sajadi \& Riazi(2017)}]{Sajadi:2017glu}
Sajadi, S.~N., \& Riazi, N. 2017, Gen. Rel. Grav., 49, 45,
  \dodoi{10.1007/s10714-017-2209-8}

\bibitem[{Sakharov(1966)}]{Sakharov:1966aja}
Sakharov, A.~D. 1966, Zh. Eksp. Teor. Fiz., 49, 345

\bibitem[{Schee \& Stuchlik(2015)}]{Schee:2015nua}
Schee, J., \& Stuchlik, Z. 2015, JCAP, 1506, 048,
  \dodoi{10.1088/1475-7516/2015/06/048}

\bibitem[{Shaikh(2019)}]{Shaikh:2019fpu}
Shaikh, R. 2019, Phys. Rev., D100, 024028, \dodoi{10.1103/PhysRevD.100.024028}

\bibitem[{Simpson \& Visser(2020)}]{Simpson:2019mud}
Simpson, A., \& Visser, M. 2020, Universe, 6, 8

\bibitem[{Stuchlik \& Schee(2014)}]{Stuchlik:2014qja}
Stuchlik, Z., \& Schee, J. 2014, Int. J. Mod. Phys., D24, 1550020,
  \dodoi{10.1142/S0218271815500200}

\bibitem[{Teo(2003)}]{Teo:2003}
Teo, E. 2003, General Relativity and Gravitation, 35, 1909

\bibitem[{Toshmatov {et~al.}(2014)Toshmatov, Ahmedov, Abdujabbarov, \&
  Stuchlik}]{Toshmatov:2014nya}
Toshmatov, B., Ahmedov, B., Abdujabbarov, A., \& Stuchlik, Z. 2014, Phys. Rev.,
  D89, 104017, \dodoi{10.1103/PhysRevD.89.104017}

\bibitem[{Toshmatov {et~al.}(2018)Toshmatov, Stuchlik, \&
  Ahmedov}]{Toshmatov:2018cks}
Toshmatov, B., Stuchlik, Z., \& Ahmedov, B. 2018, Phys. Rev., D98, 028501,
  \dodoi{10.1103/PhysRevD.98.028501}

\bibitem[{Tsukamoto {et~al.}(2014)Tsukamoto, Li, \& Bambi}]{Tsukamoto:2014tja}
Tsukamoto, N., Li, Z., \& Bambi, C. 2014, JCAP, 1406, 043,
  \dodoi{10.1088/1475-7516/2014/06/043}

\bibitem[{Tsupko(2017)}]{Tsupko:2017rdo}
Tsupko, O.~{\relax Yu}. 2017, Phys. Rev., D95, 104058,
  \dodoi{10.1103/PhysRevD.95.104058}

\bibitem[{Vagnozzi \& Visinelli(2019)}]{Vagnozzi:2019apd}
Vagnozzi, S., \& Visinelli, L. 2019, Phys. Rev., D100, 024020,
  \dodoi{10.1103/PhysRevD.100.024020}

\bibitem[{Wang {et~al.}(2019)Wang, Xu, \& Wei}]{Wang:2018prk}
Wang, H.-M., Xu, Y.-M., \& Wei, S.-W. 2019, JCAP, 1903, 046,
  \dodoi{10.1088/1475-7516/2019/03/046}

\bibitem[{Wang {et~al.}(2018)Wang, Chen, \& Jing}]{Wang:2018eui}
Wang, M., Chen, S., \& Jing, J. 2018, Phys. Rev., D98, 104040,
  \dodoi{10.1103/PhysRevD.98.104040}

\bibitem[{Will(2006)}]{Will:2005va}
Will, C.~M. 2006, Living Rev. Rel., 9, 3, \dodoi{10.12942/lrr-2006-3}

\bibitem[{Yumoto {et~al.}(2012)Yumoto, Nitta, Chiba, \&
  Sugiyama}]{Yumoto:2012kz}
Yumoto, A., Nitta, D., Chiba, T., \& Sugiyama, N. 2012, Phys. Rev., D86,
  103001, \dodoi{10.1103/PhysRevD.86.103001}

\end{thebibliography}
\bibliographystyle{aasjournal}
\end{document}